\documentclass[letterpaper,twocolumn,prl,aps,superscriptaddress,amsmath,amssymb,floatfix]{revtex4}
\usepackage{mathptmx}
\usepackage[latin9]{inputenc}
\usepackage{color}
\usepackage{verbatim}
\usepackage{float}
\usepackage{amsmath}
\usepackage{amssymb}
\usepackage{graphicx}
\usepackage{esint}

\usepackage[unicode=true,
 bookmarks=true,bookmarksnumbered=false,bookmarksopen=false,
 breaklinks=false,pdfborder={0 0 1},backref=false,colorlinks=true]
 {hyperref}
\hypersetup{
 linkcolor=magenta,urlcolor=blue,citecolor=blue,pdfstartview={FitH},hyperfootnotes=false}

\makeatletter

\usepackage{textcomp}
\usepackage{epstopdf}

\pdfpageheight\paperheight
\pdfpagewidth\paperwidth

\@ifundefined{textcolor}{}{%
 \definecolor{BLACK}{gray}{0}
 \definecolor{WHITE}{gray}{1}
 \definecolor{RED}{rgb}{1,0,0}
 \definecolor{GREEN}{rgb}{0,1,0}
 \definecolor{BLUE}{rgb}{0,0,1}
 \definecolor{CYAN}{cmyk}{1,0,0,0}
 \definecolor{MAGENTA}{cmyk}{0,1,0,0}
 \definecolor{YELLOW}{cmyk}{0,0,1,0}
}

\usepackage{xcolor}
\usepackage{soul}
\definecolor{blue}{rgb}{0,0,1}
\definecolor{red}{rgb}{1,0,0}
\definecolor{green}{rgb}{0,1,0}

\usepackage{soul}

\makeatother

\begin{document}
\title{Autonomous quantum error correction beyond break-even and its metrological application}

\author{Zhongchu Ni}
\thanks{These authors contributed equally to this work.}
\affiliation{International Quantum Academy, Shenzhen 518048, China}

\author{Ling Hu}
\thanks{These authors contributed equally to this work.}
\affiliation{International Quantum Academy, Shenzhen 518048, China}

\author{Yanyan Cai}
\thanks{These authors contributed equally to this work.}
\affiliation{International Quantum Academy, Shenzhen 518048, China}
\affiliation{Shenzhen Institute for Quantum Science and Engineering, Southern University of Science and Technology, Shenzhen 518055, China}

\author{Libo Zhang}
\author{Jiasheng Mai}
\affiliation{International Quantum Academy, Shenzhen 518048, China}
\affiliation{Shenzhen Institute for Quantum Science and Engineering, Southern University of Science and Technology, Shenzhen 518055, China}

\author{Xiaowei Deng}
\author{Pan Zheng}
\affiliation{International Quantum Academy, Shenzhen 518048, China}

\author{Song Liu}
\affiliation{International Quantum Academy, Shenzhen 518048, China}
\affiliation{Shenzhen Branch, Hefei National Laboratory, Shenzhen 518048, China}

\author{Shi-Biao Zheng}
\email{t96034@fzu.edu.cn}
\affiliation{Fujian Key Laboratory of Quantum Information and Quantum Optics, College of Physics and Information Engineering, Fuzhou University, Fuzhou 350108, China}
\affiliation{Hefei National Laboratory, Hefei 230088, China}

\author{Yuan Xu}
\email{xuyuan@iqasz.cn}
\affiliation{International Quantum Academy, Shenzhen 518048, China}
\affiliation{Shenzhen Branch, Hefei National Laboratory, Shenzhen 518048, China}

\author{Dapeng Yu}
\affiliation{International Quantum Academy, Shenzhen 518048, China}
\affiliation{Shenzhen Branch, Hefei National Laboratory, Shenzhen 518048, China}

\begin{abstract}
\textbf{The ability to extend the lifetime of a logical qubit beyond that of the best physical qubit available within the same system, i.e., the break-even point, is a prerequisite for building practical quantum computers. So far, this point has been exceeded through active quantum error correction (QEC) protocols, where a logical error is corrected by measuring its syndrome and then performing an adaptive correcting operation. Autonomous QEC (AQEC), without the need for such resource-consuming measurement-feedback control, has been demonstrated in several experiments, but none of which has unambiguously reached the break-even point. Here, we present an unambiguous demonstration of beyond-break-even AQEC in a circuit quantum electrodynamics system, where a photonic logical qubit encoded in a superconducting microwave cavity is protected against photon loss through autonomous error correction, enabled by engineered dissipation. Under the AQEC protection, the logical qubit achieves a lifetime surpassing that of the best physical qubit available in the system by 18\%. We further employ this AQEC protocol to enhance the precision for measuring a slight frequency shift, achieving a metrological gain of 6.3 dB over that using the most robust Fock-state superposition. These results illustrate that the demonstrated AQEC procedure not only represents a crucial step towards fault-tolerant quantum computation but also offers advantages for building robust quantum sensors.
}
\end{abstract}
\maketitle
\vskip 0.5cm

Quantum bits (qubits) differ from classical bits by the existence of quantum coherence between the computational basis states---a fundamental property that enables quantum computers to solve problems intractable for classical systems. However, this coherence is highly vulnerable to environmental noise, which can cause decoherence and loss of quantum information. To combat the decoherence-induced errors, various quantum error correction (QEC) protocols have been developed~\cite{nielsen2010,terhal2015,campbell2017}, in which a logical qubit is encoded within a large Hilbert space, using either multiple physical qubits~\cite{shor1995,steane1996,fowler2012,ryan2021,takeda2022,krinner2022,zhao2022,putterman2025,google2025} or a single high-dimensional system, such as a bosonic mode~\cite{ofek2016,sivak2023,brock2025,ni2023,hu2019,michael2016,fluhmann2019,cai2021,joshi2021}. This redundant encoding allows error syndromes to be measured in a quantum non-demolition manner~\cite{sun2014}, thereby enabling error correction via feedback control in real time.

In recent years, experimental demonstrations of real-time measurement-based QEC have successfully extended the lifetime of a logical qubit beyond that of the best physical qubit available in the same system---surpassing what is known as the break-even point~\cite{ofek2016,ni2023,sivak2023,google2025,brock2025}. These implementations rely heavily on high-efficiency error syndrome measurements and real-time adaptive recovery operations. Such measurement-feedback control requires additional hardware and may introduce some uncorrectable errors~\cite{rosenblum2018}, which are unfavorable for the implementation of practical large-scale quantum computation.

Autonomous QEC (AQEC) has emerged as a promising alternative that is measurement-free and hardware-efficient. In AQEC, errors are coherently corrected with an ancilla, whose entropy is autonomously removed through engineered dissipation~\cite{harrington2022}. This approach eliminates the need for active syndrome measurement and real-time classical feedback, thereby avoiding measurement-introduced errors. As a result, AQEC has recently garnered significant theoretical and experimental attention~\cite{schindler2011,reed2012,waldherr2014,ma2020,gertler2021,lachance2024,li2024autonomous,debry2025,liyi2025,kerckhoff2010,kapit2016,lihm2018,lebreuilly2021,leghtas2013,kwon2022,zeng2023,xu2023,shtanko2025}. In circuit quantum electrodynamics (QED) systems~\cite{blais2021}, proof-of-principle demonstrations of AQEC have been reported using multiphoton bosonic encodings~\cite{ma2020,gertler2021,lachance2024}, but the lifetime of the protected logical qubit is still shorter than that of the best physical qubit, which is encoded in the two lowest Fock states. Recent AQEC experiments performed in a trapped ion system have demonstrated the protection of a logical qubit encoded in the Zeeman sublevels of the metastable state of a $^{40}\mathrm{Ca}^+$ ion against dephasing errors~\cite{debry2025,liyi2025}. In these experiments, the physical qubit used for the QEC performance benchmarking was encoded in the sublevels of either the ion's ground state~\cite{debry2025} or metastable state~\cite{liyi2025}, whereas the most dephasing-insensitive physical qubit in this system is encoded between the ground state and the metastable state~\cite{schindler2013}.

\begin{figure*}
\centering
\includegraphics{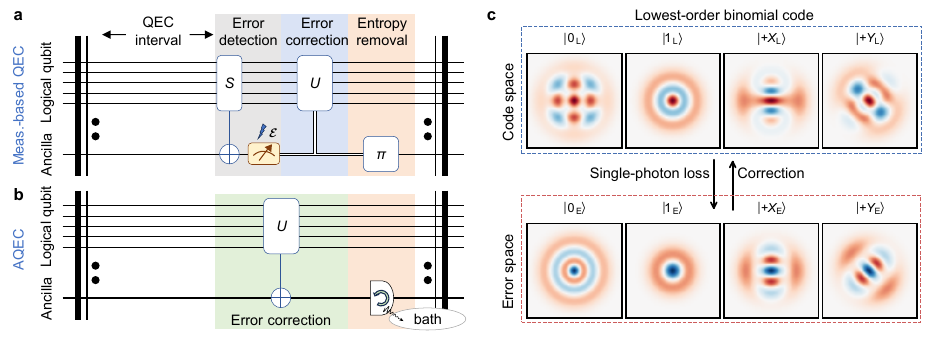} \caption{\textbf{Schematic comparison of the measurement-based QEC and AQEC procedures, and the codewords used in the experiment.} 
\textbf{a}, Conventional measurement-based QEC procedure, with each cycle consisting of three stages: error detection with ancilla measurement, real-time error correction based on the error syndrome measurements, and entropy removal to reset ancilla via classical feedback. This measurement-based QEC risks propagating ancilla measurement errors to the logical qubit. 
\textbf{b}, AQEC employs a coherent controlled-unitary operation $\mathcal{R}=|\psi_\mathrm{L}\rangle\langle \psi_\mathrm{L}|\otimes |g\rangle \langle g| + |\psi_\mathrm{L}\rangle\langle \psi_\mathrm{E}|\otimes |e\rangle \langle g|$ to perform autonomous error correction without ancilla measurement, and the ancilla is then reset through engineered dissipation to remove the error entropy to a lossy bath. Here $|\psi_\mathrm{L/E}\rangle$ is the code/error state of the logical qubit and $|g/e\rangle$ is the ancilla ground/excited state.
\textbf{c}, Wigner function representations of the lowest-order binomial codewords \{$|0_\mathrm{L}\rangle = (|0\rangle + |4\rangle)/\sqrt{2}$ and $|1_\mathrm{L}\rangle = |2\rangle$\}. The dominant single-photon loss converts the logical state from code space into the error space, spanned by \{$|0_{\mathrm{E}}\rangle = |3\rangle$, $|1_{\mathrm{E}}\rangle = |1\rangle$\}. The cardinal states on the Bloch spheres within the code and error spaces are defined as $|+X_{\mathrm{L}(\mathrm{E})}\rangle = (|0_{\mathrm{L}(\mathrm{E})}\rangle + |1_{\mathrm{L}(\mathrm{E})}\rangle)/\sqrt{2}$ and $|+Y_{\mathrm{L}(\mathrm{E})}\rangle = (|0_{\mathrm{L}(\mathrm{E})}\rangle + i|1_{\mathrm{L}(\mathrm{E})}\rangle)/\sqrt{2}$.
}
\label{fig1}
\end{figure*}

Here, we report an unambiguous demonstration of AQEC that extends the lifetime of a logical qubit beyond the break-even point in a circuit QED system. In this experiment, the logical qubit is binomially encoded in the photonic mode stored in a high-quality superconducting microwave cavity, which is dispersively coupled to a transmon qubit that serves as an ancilla. The environmentally-induced error entropy is coherently transferred to the ancilla qubit, and then removed through engineered dissipation with the assistance of a leaky resonator. This error correction procedure autonomously brings the logical qubit back to its original codeword state once a photon has been lost. 
Under such AQEC protection, the encoded logical qubit has a lifetime longer than that of the best available physical qubit by 18\%. 
In addition, we incorporate this AQEC procedure into a quantum metrological experiment for sensing the frequency shift of the photonic mode, achieving a maximum Fisher information gain of 6.3 dB over the uncorrected physical qubit encoded with the two lowest Fock states.
Our results open a promising perspective of AQEC for implementing fault-tolerant quantum computation, as well as for improving the performance of quantum sensors.

Figure~\ref{fig1}a illustrates the conventional measurement-based QEC procedure. Each QEC cycle consists of three main stages: error detection, implemented via a stabilizer-controlled-NOT operation that maps the error syndrome onto an ancilla qubit, which is then measured; error correction, which restores the logical state by applying a conditional recovery operation based on the measurement outcomes through classical feedback in real time; and entropy removal, where the ancilla is reset to its ground state through measurement-feedback control to permit reuse in subsequent cycles. These three steps are executed repeatedly with a fixed cycle interval. 
This approach, however, suffers from inherent limitations such as ancilla classical hardware overhead and measurement error propagation.

\begin{figure*}
\includegraphics{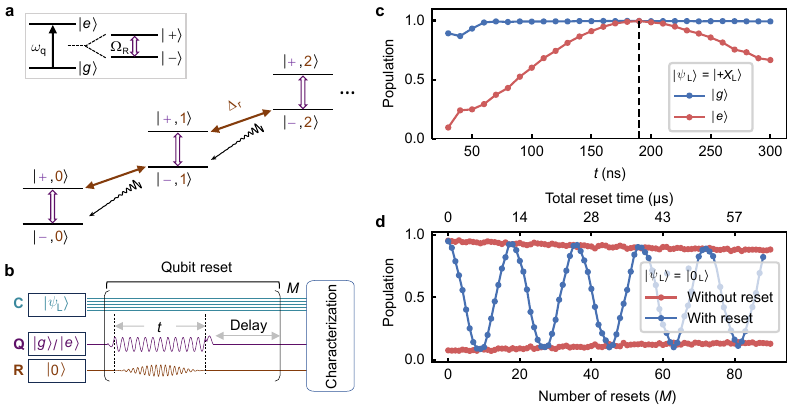} \caption{\textbf{Ancilla qubit reset via reservoir engineering.} 
\textbf{a}, Energy level diagram of the system combined with the ancilla and a lossy resonator, which acts as a bath to dump the error entropy into the environment. A resonant drive applied to the ancilla with a Rabi frequency $\Omega_{\mathrm{R}}$ generates dressed qubit states $|\pm\rangle$ with an energy splitting $\Omega_{\mathrm{R}}$ (inset). A sideband drive applied to the resonator with a detuning $\Delta_{\mathrm{r}}=\Omega_{\mathrm{R}}$ induces transition from $|+,n\rangle$ to $|-,n+1\rangle$ (brown arrows), followed by rapid relaxation of the resonator (black arrows), resetting the system to $|-,0\rangle$. 
\textbf{b}, Experimental sequence for qubit reset characterization. The cavity is prepared in the logical state $|\psi_{\mathrm{L}}\rangle$ and the ancilla in $|g\rangle$ or $|e\rangle$. After applying $M$ reset cycles---each consisting of simultaneous qubit and resonator drives, optimized qubit pulses for basis transformation, and a resonator relaxation delay---the logical or ancilla qubit states are measured through a characterization process. 
\textbf{c}, Measured ancilla ground-state population versus reset drives' duration $t$ for one reset cycle ($M=1$) with an initial logical state $|+X_{\mathrm{L}}\rangle$. An average ground state probability of $\sim 99.6\%$ is achieved at $t=190$~ns regardless of the initial ancilla states. 
\textbf{d}, Measured phase coherence of the logical qubit state $|0_{\mathrm{L}}\rangle$ as a function of the number of qubit resets (blue) and the total idling time without reset (red).
  }
\label{fig2} \vspace{-6pt}
\end{figure*}

By contrast, the AQEC protocol realizes error correction through a coherent controlled-unitary operation between the ancilla and logical qubit system, as illustrated in Fig.~\ref{fig1}b. This procedure coherently and autonomously brings the logical qubit from the error space back to the code space in a fully quantum manner, and the ancilla is then reset via engineered dissipation, which removes the error entropy from the ancilla to the environment. This AQEC approach circumvents the need for ancilla measurement and classical feedback, thereby making the procedure resource-efficient, which is favorable for implementing scalable quantum computation. 

We here incorporate a high-efficiency autonomous ancilla-reset technique~\cite{murch2012} with the photon-loss-correcting procedure developed in Ref.~\cite{ma2020}, implementing a beyond-break-even AQEC procedure in a circuit QED system. 
The logical qubit is encoded in the photonic mode stored in a high-quality microwave cavity with the lowest-order binomial codewords~\cite{michael2016,hu2019,ni2023},
\begin{eqnarray}
|0_\mathrm{L}\rangle = \frac{|0\rangle + |4\rangle}{\sqrt{2}}, \quad |1_\mathrm{L}\rangle = |2\rangle,
\end{eqnarray}
where the number in each ket denotes the photon number of the corresponding Fock states. The dominant error channel---single-photon loss---makes the logical qubit jump to the error subspace spanned by \{$|0_\mathrm{E}\rangle = |3\rangle,|1_\mathrm{E}\rangle = |1\rangle$\}. Figure~\ref{fig1}c visualizes these logical and error states via their Wigner functions in phase space. The stabilizer operator for this code is the photon number parity $S=\exp{(i\pi a^\dagger a)}$, as the logical code-space contains only even photon-number states, while error-space states exhibit odd parity. Here, $a^\dagger$ and $a$ denote the creation and annihilation operators for the cavity mode.

In our experimental setup, the microwave cavity $C$ (with single-photon relaxation time \(T_{1}^\mathrm{C} = 1.4~\mathrm{ms}\) and pure dephasing time \(T_{\phi}^\mathrm{C} = 6.5~\mathrm{ms}\)) is dispersively coupled to a superconducting transmon qubit $Q$ (with coherence times \(T_{1}^\mathrm{q} = 0.13~\mathrm{ms}\) and \(T_{\phi}^\mathrm{q} = 0.39~\mathrm{ms}\)) that serves as the ancilla for AQEC. The dispersive cavity-ancilla interaction is described by Hamiltonian \(H_\mathrm{qc}=-\chi_\mathrm{qc} a^\dagger a |e\rangle\langle e|\) (assuming $\hbar=1$), where $|e\rangle$ ($|g\rangle$) denotes the excited (ground) state of the ancilla and $\chi_\mathrm{qc}=2\pi\times 0.88$~MHz is the dispersive shift. We employ quantum optimal control (QOC) pulses designed via the gradient-accent pulse engineering (GRAPE) technique~\cite{khaneja2005,heeres2017} to implement the unitary error correcting operation $\mathcal{R}=|\psi_\mathrm{L}\rangle\langle \psi_\mathrm{L}|\otimes |g\rangle \langle g| + |\psi_\mathrm{L}\rangle\langle \psi_\mathrm{E}|\otimes |e\rangle \langle g|$ that performs the state transfer~\cite{ma2020}:
\begin{eqnarray}
\left(|\psi_{\mathrm{L}}\rangle \oplus |\psi_{\mathrm{E}}\rangle \right) |g\rangle \xrightarrow{\mathcal{R}} |\psi_{\mathrm{L}}\rangle \left(|g\rangle \oplus |e\rangle \right),
\end{eqnarray}   
where $|\psi_{\mathrm{L}/\mathrm{E}}\rangle = \alpha|0_{\mathrm{L}/\mathrm{E}}\rangle+\beta|1_{\mathrm{L}/\mathrm{E}}\rangle$ ($|\alpha|^2 +|\beta|^2=1$) represents an arbitrary state in the logical or error subspace with $S|\psi_{\mathrm{L}/\mathrm{E}}\rangle = \pm |\psi_{\mathrm{L}/\mathrm{E}}\rangle$. Note that the recovery operation of the GRAPE pulse also compensates for the no-jump backaction effect on the logical state associated with no photon loss (see SM).

\begin{figure*}
\includegraphics{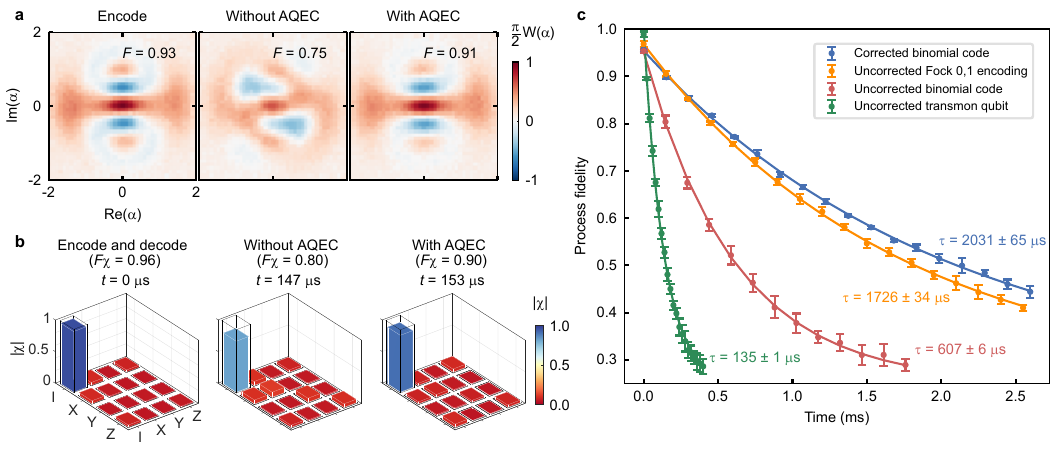} \caption{\textbf{Performance of the AQEC procedure.} 
\textbf{a}, Measured Wigner functions of the logical state $|+X_{\mathrm{L}}\rangle$: immediately after encoding (left), after idling $\sim150\mu$s without AQEC (middle), and after a single AQEC cycle with a duration $\sim150\mu$s (right). 
\textbf{b}, Bar charts of the magnitudes of the quantum process $\chi$ matrices for three cases: encode/decode only (left); interleaved with an idling process without AQEC (middle); and interleaved with a single AQEC cycle (right). 
\textbf{c}, Measured process fidelity as a function of storage time for different encodings. Error bars represent one standard deviation obtained from repeated experiments. All curves are fitted to $F_{\chi} = A e^{-t/\tau} + 0.25$ to extract the lifetime $\tau$ of the corresponding encodings, with uncertainties obtained from the fits. The extracted $\tau$ of the corrected binomial code (blue) surpasses the uncorrected binomial code (red) and uncorrected transmon qubit (green), and even exceeds the break-even point (orange) by a factor of 18\%. 
}
\label{fig3} \vspace{-6pt}
\end{figure*}

Following the recovery process $\mathcal{R}$, which simultaneously corrects the error state and transfers error entropy to the ancilla qubit, it is necessary to reset the ancilla for reuse in subsequent QEC cycles. In our experiment, the ancilla is reset through a quantum bath engineering technique~\cite{murch2012}, realized by dispersively coupling the ancilla to its readout resonator $R$ (with a decay rate $\kappa_{\mathrm{r}}=2\pi\times 1.4$~MHz), which serves as a lossy channel to transfer the entropy from the ancilla to the environment. Under the strong dispersive qubit-resonator interaction $\chi_\mathrm{qr}=2\pi\times 3.0$~MHz, we apply a resonant Rabi drive to the ancilla, generating dressed qubit states $|\pm\rangle = (|g\rangle \pm |e\rangle)/\sqrt{2}$ with an energy splitting of $\Omega_{\mathrm{R}}$ (the Rabi frequency of the drive), as illustrated in the inset of Fig.~\ref{fig2}a. Concurrently, an off-resonant drive applied to the resonator with a detuning $\Delta_{\mathrm{r}} = \Omega_{\mathrm{R}}$  mediates the sideband transition $|+,n\rangle \leftrightarrow |-,n+1\rangle$, with the energy level diagram of the joint qubit-resonator states illustrated in Fig.~\ref{fig2}a. Here $n$ denotes the photon number in the resonator. Due to the fast spontaneous decay of the lossy resonator, this process irreversibly resets the joint system to $|-,0\rangle$, regardless of the initial qubit states. 

The experimental sequence for characterizing this qubit reset process is presented in Fig.~\ref{fig2}b. After preparing a logical state $|+X_{\mathrm{L}}\rangle = (|0_{\mathrm{L}}\rangle + |1_{\mathrm{L}}\rangle)/\sqrt{2}$, we measure the ancilla ground-state population versus the duration of the reset drives $t$, with the ancilla starting from $|g\rangle$ or $|e\rangle$. Here, additional optimized qubit pulses are applied before and after the qubit Rabi drive for basis transformation, ensuring the ancilla is reset to the $|g\rangle$ state (see SM). Figure~\ref{fig2}c shows the measured ground-state populations as a function of $t$ for different initial ancilla states. With a reset time of about $t=190$~ns, the process achieves an unconditional qubit reset with an average ground state probability of 99.6\%, independent of the initial ancilla state and the cavity's photon number due to a large drive amplitude $\Omega_{\mathrm{R}}\gg\chi_\mathrm{qc}$. To further verify that the reset process introduces negligible dephasing errors on the logical states within the cavity, we measure the phase coherence of the logical state $|0_{\mathrm{L}}\rangle$ under repeated resets and compare it with free idling evolution (Fig.~\ref{fig2}d). The similar exponential decay rates of these curves confirm that each reset operation induces minimal dephasing errors on logical states.

With these key components characterized, we proceed to demonstrate autonomous protection of the binomially-encoded logical qubit against single-photon loss with the AQEC. After encoding the logical state $|+X_{\mathrm{L}}\rangle$, we measure the cavity Wigner functions with and without applying the AQEC cycle (duration $\sim$150~$\mu$s), as shown in Fig.~\ref{fig3}a. The results confirm that the AQEC well protects the logical qubit from single-photon loss. We further quantify the AQEC performance through measuring the quantum process $\chi$ matrix for one AQEC cycle, compared to that without AQEC protection (Fig~\ref{fig3}b). The improved process fidelity further verifies the effectiveness of the AQEC. 

The key metric for quantifying the AQEC performance is the extension of the logical qubit lifetime relative to the best physical qubit available in the system. In our 3D circuit QED device, the best physical qubit is encoded in the two lowest Fock states $|0\rangle$ and $|1\rangle$, which, referred to as the Fock qubit, is most robust against dissipation without QEC protection. Figure~\ref{fig3}c shows the measured process fidelity as a function of the storage time with the binomial code under AQEC protection (blue), compared to the unprotected binomial code (red), transmon qubit (green), and Fock qubit (orange). We determine the lifetimes $\tau$ by fitting all curves to the function $F_{\chi}(t) = A e^{-t/\tau} + 0.25$. As a result, the AQEC-protected binomial code demonstrates a lifetime of $\sim2.0$~ms, which is 15.0 times longer than that of the transmon qubit and 3.3 times longer than the unprotected binomial code. Most notably, this lifetime is 1.18 times longer than that of the best physical qubit in this system.

\begin{figure*}
\includegraphics{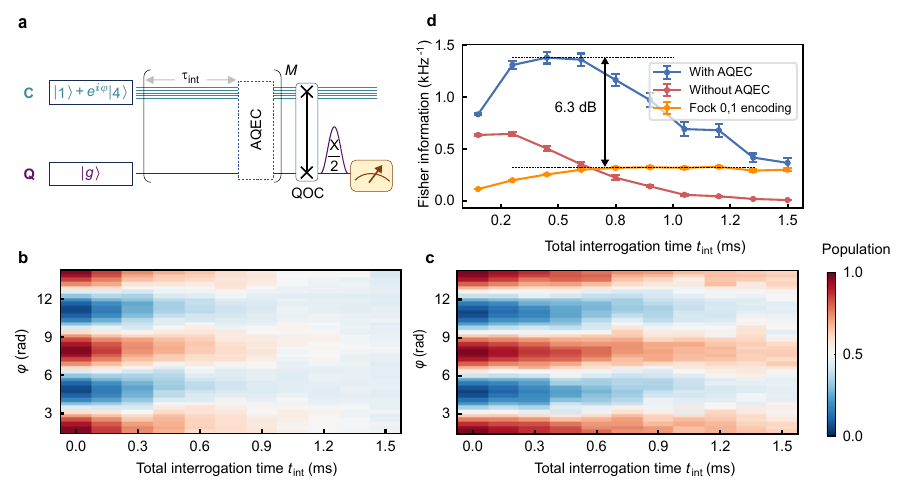} \caption{\textbf{Ramsey interference under AQEC protection for quantum sensing.} 
\textbf{a}, Experimental sequence for Ramsey interferometry experiments with and without AQEC protection using superposition of Fock states. 
\textbf{b-c}, Measured ancilla ground-state population oscillating as a function of the initial phase $\varphi$ with different total interrogation times $t_\mathrm{int}=M\tau_\mathrm{int}$ for the cases without (b) and with (c) AQEC protection. Each AQEC cycle has a fixed duration of $\tau_\mathrm{int}=150~\mu$s.  
\textbf{d} Extracted normalized Fisher information from (b-c) as a function of the total interrogation time with (blue) and without (red) AQEC protection, as well as that using Fock 0,1, encoding (orange). Error bars are obtained from the error propagation of the fitting parameter uncertainties. A maximum Fisher information gain of 6.3 dB is achieved over the uncorrected Fock 0,1 encoding.  
}
\label{fig4} \vspace{-6pt}
\end{figure*}

The demonstrated AQEC protocol, which extends the logical lifetime beyond the break-even point, can be further applied to quantum metrology to improve the estimation precision~\cite{dur2014,arrad2014,kessler2014,unden2016,reiter2017} of a slight frequency shift, described by the Hamiltonian $H_\mathrm{int}=\delta a^\dagger a$. Although exact error-correcting codes are excluded from achieving the Heisenberg limit for sensing $\delta$ due to the non-satisfaction of the "Hamiltonian-not-in-Lindblad span" condition~\cite{zhou2018,wang2022NC}, an approximate QEC code can still benefit quantum metrology~\cite{wang2022NC,zhou2020}. In the experiment reported in Ref.~\cite{wang2022NC}, the metrological quantum coherence was protected by a semi-autonomous QEC procedure, where the ancilla was reset by measurement-feedback control. Here we demonstrate the metrological application of the fully autonomous, resource-efficient QEC with a Ramsey interferometer acting on the Fock superposition state $|1\rangle + e^{i\varphi}|4\rangle$ (ignoring the normalization factor) with variable phase $\varphi$ (experimental sequence shown in Fig.~\ref{fig4}a). Note that although this encoding lacks definite photon number parity as an error syndrome, our AQEC procedure remains applicable (see SM). After an interrogation time $\tau_\mathrm{int}$, the cavity state evolves into $|1\rangle + e^{i(3\delta \tau_\mathrm{int} + \varphi)}|4\rangle$. Although single-photon loss transforms this superposition state into the error state $|0\rangle+2e^{i(3\delta \tau_\mathrm{int} + \varphi)}|3\rangle$ with deformation, the phase coherence can be effectively preserved and protected by repetitive AQEC procedures.

As shown in Figs.~\ref{fig4}b-c, we compare the measured ancilla ground-state population $P_g$, which oscillates with the initial phase $\varphi$ for different total interrogation times $t_\mathrm{int}=M\tau_\mathrm{int}$, for cases without and with AQEC protection. The Ramsey fringes are fitted to a sinusoidal function $P_g=A\cos(3\delta t_\mathrm{int}+\varphi)+B$ with fitting parameters $A$ and $B$. The results indicate that AQEC protection significantly slows the decay of Ramsey contrast, thus preserving the phase coherence between the two Fock states. To quantify the sensing performance, we compute the normalized Fisher information using the formula~\cite{wang2022NC,deng2024}
\begin{eqnarray}
F=\max_\varphi{\left(\frac{1}{P_g(1-P_g)t_\mathrm{tot}}\left(\frac{\partial P_g}{\partial \delta}\right)^2\right)} = \frac{9A^2t_\mathrm{int}^2}{B(1-B)t_\mathrm{tot}},
\end{eqnarray}
which has units of $\mathrm{Hz}^{-1}$ and determines the maximum achievable sensitivity to $\delta$. Here, $t_\mathrm{tot}$ denotes the total single-shot measurement time, including initialization, interrogation, characterization, readout, and AQEC operation time. Figure~\ref{fig4}d presents the extracted normalized Fisher information for $|1\rangle + e^{i\varphi}|4\rangle $ versus the total interrogation time with and without AQEC protection, along with that of the uncorrected $|0\rangle + e^{i\varphi}|1\rangle $ for reference (see SM). The Fisher information reaches a maximum at an optimal total interrogation time $t_\mathrm{int}$, reflecting a trade-off between phase accumulation (benefiting from longer $t_\mathrm{int}$) and decoherence (minimized with shorter $t_\mathrm{int}$). Under AQEC protection, the maximum Fisher information reaches $1.4~\mathrm{kHz}^{-1}$, surpassing the unprotected case without AQEC by 3.3~dB and the result for $|0\rangle + e^{i\varphi}|1\rangle $ by 6.3~dB. These results underscore the potential of AQEC-enhanced sensors for high-precision quantum metrology via the Ramsey interferometer.

In conclusion, we have demonstrated a fully measurement-free QEC scheme that protects a bosonic logical qubit against single-photon loss errors beyond the break-even point in a circuit QED system. By encoding quantum information in a binomial photonic qubit and using a dissipatively engineered ancilla reset process, we have realized a measurement-free and feedback-free error correction cycle that autonomously corrects photon loss errors while avoiding ancilla measurement error propagation. The logical qubit lifetime under repetitive AQEC protection exceeds that of the best physical qubit in the same system by 18\%, providing unambiguous experimental evidence that the break-even point can be surpassed without resorting to error syndrome measurement and active feedback control. This AQEC approach is versatile and applicable to various error-correcting codes, even those lacking a specific photon number parity for error syndromes. Additionally, it is applicable to correct multiple types of errors concurrently, such as multi-photon losses and dephasing errors (see SM). Furthermore, we show the potential application of this AQEC strategy in quantum metrology, observing enhanced phase coherence in a Ramsey interferometer. Our hardware-efficient and measurement-free QEC approach may serve as a foundational component in the development of fault-tolerant quantum computation and robust quantum sensors, opening promising perspectives for practical applications of quantum technologies.

\textit{Note added}--After completion of our manuscript, we note a demonstration of AQEC in a similar system where the break-even point is surpassed by 4\% with the assistance of multiple ancilla qubits~\cite{sun2025}.

\smallskip{}

%

\clearpage{}
\setcounter{figure}{0} 
\noindent \textbf{\large{}Methods}{\large\par}

\noindent \textbf{Experimental device}

\noindent The AQEC experiments are performed using a three-dimensional (3D) circuit QED device~\cite{blais2021,paik2011,vlastakis2013,kirchmair2013} comprising a high-quality microwave cavity, a transmon qubit~\cite{koch2007}, and a lossy stripline resonator~\cite{axline2016}. The microwave cavity is designed with a 3D coaxial $\lambda/4$ stub cavity~\cite{reagor2016} that is machined from a high-purity aluminum block and chemically etched~\cite{reagor2013} to achieve a high quality factor of $5.8\times 10^7$. This cavity has a resonance frequency of 6.61~GHz and is used for storing the multiphoton bosonic logical qubit. The transmon qubit, with a transition frequency of 5.48~GHz and coherence times $T_1=0.13$~ms and $T_2^*=0.16$~ms, serves as an ancilla for implementing error correction autonomously, as well as the encoding and decoding processes. The stripline resonator, which is patterned on the same chip as the ancilla with a thin tantalum film~\cite{place2021, wang2022npjQI}, has a resonance frequency of 8.58~GHz and a fast decay rate of $\kappa_\mathrm{r}/2\pi=1.4$~MHz, facilitating both ancilla reset via dissipation engineering and final qubit readout. The ancilla qubit is dispersively coupled to both the 3D cavity and stripline resonator with dispersive shifts $\chi_\mathrm{qc}/2\pi=0.88$~MHz and $\chi_\mathrm{qr}/2\pi=3.0$~MHz, respectively. All control and measurement operations are implemented using a Zurich Instruments SHFQC quantum controller.

\vbox{}

\noindent \textbf{Ancilla reset via driven dissipation}

\noindent The ancilla reset required for entropy removal in the AQEC cycle is implemented by simultaneously applying a resonant Rabi drive to the ancilla and an off-resonant sideband drive to the lossy resonator~\cite{murch2012} with a large decay rate $\kappa_r$. In the rotating frame of the ancilla and resonator frequencies, the driven dispersive Hamiltonian of the joint system is given by 
\begin{eqnarray}
H_d = \Delta_\mathrm{r} a_\mathrm{r}^\dagger a_\mathrm{r} - \frac{\chi_\mathrm{qr}}{2}a_\mathrm{r}^\dagger a_\mathrm{r} \sigma_z + \frac{\Omega_\mathrm{R}}{2}\sigma_x + \epsilon_\mathrm{r} \left( a_\mathrm{r}^\dagger + a_\mathrm{r} \right),
\end{eqnarray}
where $a_\mathrm{r}^\dagger$, $a_\mathrm{r}$ are the resonator ladder operators, $\sigma_{x,y,z}$ are the ancilla Pauli operators, $\Omega_\mathrm{R}$ is the Rabi drive amplitude, and $\epsilon_\mathrm{r}$ and $\Delta_\mathrm{r}$ are the sideband drive amplitude and frequency detuning, respectively.

Applying a displaced frame transformation $D(\xi_\mathrm{r})=\exp{(\xi_\mathrm{r} a_\mathrm{r}^\dagger - \xi_\mathrm{r}^* a_\mathrm{r})}$ with $\xi_\mathrm{r}=\frac{-\varepsilon_\mathrm{r}}{\Delta_\mathrm{r} -i\kappa_\mathrm{r}/2}$ and an ancilla Hadamard rotation transformation,
leads to the effective Hamiltonian under the rotating-wave approximation:
\begin{eqnarray}
    H_\mathrm{eff} = \Delta_\mathrm{r} a_\mathrm{r}^\dagger a_\mathrm{r} + \frac{\Omega_\mathrm{R}}{2}\tilde{\sigma}_z - \frac{\chi_\mathrm{qr}}{2}\left(\xi_\mathrm{r} a_\mathrm{r}^\dagger \tilde{\sigma}_- + \xi_\mathrm{r}^* a_\mathrm{r} \tilde{\sigma}_+ \right),
\end{eqnarray}
where $\tilde{\sigma}_z$ and $\tilde{\sigma}_\pm$ are Pauli operators in the dressed basis \{$|+\rangle$, $|-\rangle$\} formed by the Rabi drive. Setting $\Delta_\mathrm{r} = \Omega_\mathrm{R}$ resonantly couples the dressed ancilla and resonator, enabling excitation exchange. This exchange interaction combined with the large resonator decay would irreversibly transfer ancilla excitations into resonator photons that rapidly leak into the environment, thereby resetting the ancilla with an effective loss rate $\chi_\mathrm{qr}^2|\xi_\mathrm{r}|^2/\kappa_\mathrm{r}$ (see SM). 

In the experiment, we choose $\epsilon_\mathrm{r} = 2\pi\times 80$~MHz and $\Omega_\mathrm{R}= 2\pi\times 40~\mathrm{MHz} \gg \chi_\mathrm{qc}$ to ensure rapid and unconditional ancilla reset, independent of the cavity's photon number (see SM).

\vbox{}

\noindent \textbf{AQEC pulse design and implementation}

\noindent The error correction unitary in the AQEC protocol is implemented using an optimal control pulse, which simultaneously restores the logical states from the error space back to the code space and transfers the error entropy to the ancilla. This pulse is designed using the GRAPE algorithm~\cite{khaneja2005,heeres2017} to realize a set of targeted quantum state transfers on the joint cavity-ancilla system. 

For the binomially-encoded logical qubit, the AQEC pulse implements a unitary operation $U_\mathrm{AQEC}$ that corrects both the single-photon-loss error and the no-photon-jump backaction error by implementing the following state transformations on the joint cavity-ancilla system~\cite{ma2020}:
\begin{eqnarray}
    U_\mathrm{AQEC}|\psi_\mathrm{L}'\rangle |g\rangle = |\psi_\mathrm{L}\rangle |g\rangle, \quad U_\mathrm{AQEC}|\psi_\mathrm{E}\rangle |g\rangle = |\psi_\mathrm{L}\rangle |e\rangle,
\end{eqnarray}
where $|\psi_\mathrm{L/E}\rangle = \alpha |0_\mathrm{L/E}\rangle + \beta |1_\mathrm{L/E}\rangle$ represents an arbitrary state in the binomially-encoded logical or error subspace with $|\alpha|^2+|\beta|^2=1$, and $|\psi_\mathrm{L}'\rangle$ denotes the distorted logical state due to no-photon-jump backaction (see SM). The optimized microwave pulses are simultaneously applied to both the cavity and ancilla to realize this unitary, thereby mapping error states back to the code space and transferring the error entropy to the ancilla. 

For the approximate QEC code with two-component Fock-state encoding in the quantum sensing experiment, the AQEC pulse performs similar state transformations~\cite{wang2022NC,cai2021PRL}:
\begin{eqnarray}
    U_\mathrm{AQEC}|\psi_\mathrm{L}\rangle |g\rangle = |\psi_\mathrm{L}\rangle |g\rangle, \quad U_\mathrm{AQEC}|\psi_\mathrm{E}\rangle |g\rangle = |\psi_\mathrm{L}'\rangle |e\rangle,
\end{eqnarray}
where $|\psi_\mathrm{L}\rangle= (|1\rangle + e^{i\varphi} |4\rangle)/\sqrt{2}$ is the initial sensing state with arbitrary phase $\varphi$, $|\psi_\mathrm{E}\rangle= (|0\rangle + 2e^{i\varphi} |3\rangle)/\sqrt{2}$ represents the error state after a single-photon loss, and $|\psi_\mathrm{L}'\rangle = (|1\rangle + 2e^{i\varphi} |4\rangle)/\sqrt{2}$ corresponds to the recovered logical state with a slight deformation. Despite this deformation, the phase coherence between the two-component Fock states is preserved, enabling quantum-enhanced sensing.  

\vbox{}
\smallskip{}

\noindent \textbf{\large{}Data availability}{\large\par}

\noindent All data generated or analysed during this study are available
within the paper and its Supplementary Information. Further source
data will be made available on reasonable request.

\smallskip{}

\noindent \textbf{\large{}Code availability}{\large\par}

\noindent The code used to solve the equations presented in the Supplementary
Information will be made available on reasonable request.

\smallskip{}

\noindent \textbf{\large{}Acknowledgment}{\large\par}

\noindent This work was supported by the National Natural Science Foundation of China (Grants No.~12422416, No.~12274198, No.~12274080, No.~12374471), the Innovation Program for Quantum Science and Technology (Grants No.~2024ZD0302300, No.~2021ZD0301703, No.~2021ZD0300200), the Guangdong Basic and Applied Basic Research Foundation (Grant No.~2024B1515020013), and the Shenzhen Science and Technology Program (Grant No.~RCYX20210706092103021, No.~RCYX20221008092907026). 

\smallskip{}

\noindent \textbf{\large{}Author contributions}{\large\par}

\noindent Y.X. supervised the experiment and conceived the idea. Z.N. performed the experiments under the supervision of Y.X. Z.N. and L.H. analyzed the experimental data. Z.N. and Y.C. carried out numerical simulations. L.Z. fabricated the superconducting qubit under the supervision of S.L. S.-B.Z. provided theoretical supports and insights. J.M., X.D., and P.Z. contributed to the experimental setup and analysis. Z.N., L.H., Y.C., S.-B.Z., and Y.X. wrote the manuscript with feedback from all authors. Y.X. and D.Y. supervised the project.

\smallskip{}

\noindent \textbf{\large{}Competing interests}{\large\par}

\noindent The authors declare no competing interests.

\smallskip{}

\noindent \textbf{\large{}Additional information}{\large\par}

\noindent \textbf{Supplementary information} The online version contains
supplementary material.

\noindent \textbf{Correspondence and requests for materials} should
be addressed to S.-B.Z. and Y.X.

\clearpage{}

\end{document}


\title{Supplementary Information for ``Autonomous quantum error correction beyond break-even and its metrological application"}

\author{Zhongchu Ni}
\thanks{These authors contributed equally to this work.}
\affiliation{International Quantum Academy, Shenzhen 518048, China}

\author{Ling Hu}
\thanks{These authors contributed equally to this work.}
\affiliation{International Quantum Academy, Shenzhen 518048, China}

\author{Yanyan Cai}
\thanks{These authors contributed equally to this work.}
\affiliation{International Quantum Academy, Shenzhen 518048, China}
\affiliation{Shenzhen Institute for Quantum Science and Engineering, Southern University of Science and Technology, Shenzhen 518055, China}

\author{Libo Zhang}
\author{Jiasheng Mai}
\affiliation{International Quantum Academy, Shenzhen 518048, China}
\affiliation{Shenzhen Institute for Quantum Science and Engineering, Southern University of Science and Technology, Shenzhen 518055, China}

\author{Xiaowei Deng}
\author{Pan Zheng}
\affiliation{International Quantum Academy, Shenzhen 518048, China}

\author{Song Liu}
\affiliation{International Quantum Academy, Shenzhen 518048, China}
\affiliation{Shenzhen Branch, Hefei National Laboratory, Shenzhen 518048, China}

\author{Shi-Biao Zheng}
\email{t96034@fzu.edu.cn}
\affiliation{Fujian Key Laboratory of Quantum Information and Quantum Optics, College of Physics and Information Engineering, Fuzhou University, Fuzhou 350108, China}
\affiliation{Hefei National Laboratory, Hefei 230088, China}

\author{Yuan Xu}
\email{xuyuan@iqasz.cn}
\affiliation{International Quantum Academy, Shenzhen 518048, China}
\affiliation{Shenzhen Branch, Hefei National Laboratory, Shenzhen 518048, China}

\author{Dapeng Yu}
\affiliation{International Quantum Academy, Shenzhen 518048, China}
\affiliation{Shenzhen Branch, Hefei National Laboratory, Shenzhen 518048, China}

\maketitle
\tableofcontents

\clearpage

\section{Experimental device and setup}

\subsection{Experimental device}
\begin{figure}[b]
    \includegraphics{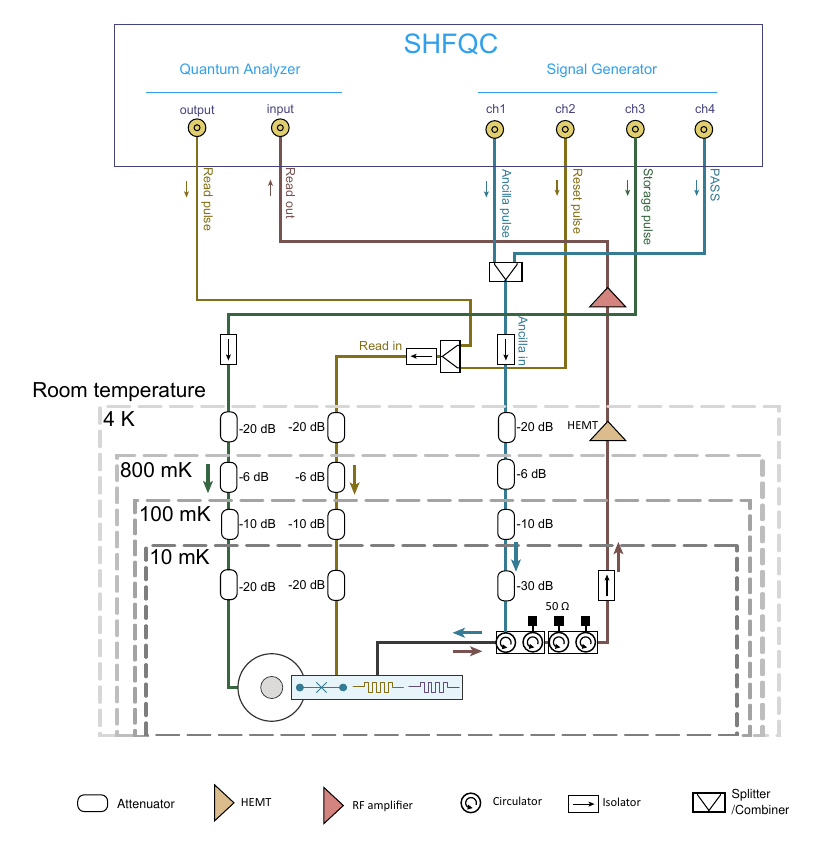}
    \caption{\textbf{Measurement setup and experimental device schematic.}} 
    \label{fig:Fig_setup}
\end{figure}

The autonomous quantum error correction (AQEC) experiment is implemented in a three-dimensional (3D) circuit quantum electrodynamics (QED) architecture~\cite{blais2021,paik2011,vlastakis2013,kirchmair2013}. The device, schematically depicted in Fig.~\ref{fig:Fig_setup}, consists of three key components: a 3D coaxial stub microwave cavity~\cite{reagor2016} for storing long-lived bosonic states, a high-coherence superconducting transmon ancilla qubit~\cite{koch2007}, and a Purcell-filtered stripline lossy resonator~\cite{axline2016,chou2018}.

The 3D microwave cavity is machined from high-purity (5N5) aluminum and chemically etched to improve coherence~\cite{reagor2013}. Its fundamental mode serves as the bosonic mode for storing microwave photons to encode the bosonic logical qubit. The transmon ancilla qubit, patterned on a sapphire chip, is dispersively coupled to the storage cavity and facilitates entropy extraction and error correction. The Josephson junction of the qubit is fabricated using a standard double-angle evaporation technique, while the qubit antenna pads and striplines are deposited with tantalum films to enhance the transmon coherence~\cite{place2021,wang2022b}.

The lossy resonator is implemented as a quasi-planar $\lambda/2$ stripline enclosed by the metal wall of its surrounding tunnel~\cite{axline2016,wang2016}. Strongly coupled to the transmon qubit, it enables fast dispersive readout and engineered dissipative reset of the ancilla. To suppress radiative decay, the resonator is coupled to an additional Purcell filter~\cite{axline2016,chou2018}. This configuration preserves the coherence of the cavity mode while allowing high-fidelity control, measurement, and reset of the ancilla qubit, which are essential for implementing repetitive AQEC cycles on the bosonic logical qubit.

\subsection{Measurement setup}

The experimental device is enclosed in a magnetic shield and installed inside a dilution refrigerator operating at a base temperature below 10~mK. A schematic of the complete experimental wiring is presented in Fig.~\ref{fig:Fig_setup}. Both the transmon ancilla qubit and the bosonic logical qubit encoded in the storage cavity are manipulated using microwave pulses synthesized by the Signal Generator channels of the Zurich Instruments SHFQC controller. Qubit control pulses have a cosine-shaped envelope with a duration of $\sim 40$~ns, incorporating the derivative-removal-by-adiabatic-gate (DRAG) technique to suppress leakage into higher excited states~\cite{Motzoi2009,Gambetta2011}. Cavity control pulses, employed for encoding and decoding the bosonic logical qubit as well as for executing AQEC operations, are obtained through numerical optimization based on the gradient ascent pulse engineering (GRAPE) method~\cite{Khaneja2005}. The reset drive applied to the resonator is also generated from the Signal Generator channels of the SHFQC. In addition, the qubit control pulses are combined with those to engineer photon-number-resolved a.c.~Stark shift (PASS) ~\cite{ma2020} before being sent to the device.

The readout pulse, generated from the Quantum Analyzer channel of the SHFQC and combined with the reset drive, is delivered to the resonator through attenuated and filtered microwave coaxial lines. The transmitted readout signal is amplified using a high-electron-mobility transistor (HEMT) amplifier at the 4~K stage, and a commercial low-noise RF amplifier at room temperature before being processed by SHFQC for ancilla state discrimination.

\subsection{System Hamiltonian and device parameters}
In the dispersive regime, the system Hamiltonian of the experimental device, comprising a storage cavity ($C$), an auxiliary transmon qubit ($Q$), and a reservoir resonator ($R$), is given by (assuming $\hbar=1$):
\begin{eqnarray}
\label{H1}
\hat{H} &=& \omega_\mathrm{q} \hat{a}_\mathrm{q}^\dagger \hat{a}_\mathrm{q} + \omega_\mathrm{c} \hat{a}_\mathrm{c}^\dagger \hat{a}_\mathrm{c} +\omega_\mathrm{r} \hat{a}_\mathrm{r}^\dagger \hat{a}_\mathrm{r} \notag\\
&-& \frac{K_\mathrm{q}}{2}\hat{a}_\mathrm{q}^{\dagger 2}\hat{a}_\mathrm{q}^{2}-\frac{K_\mathrm{c}}{2}\hat{a}_\mathrm{c}^{\dagger 2}\hat{a}_\mathrm{c}^{2} \notag\\
&-& \chi_\mathrm{qr}\hat{a}_\mathrm{q}^\dagger\hat{a}_\mathrm{q}\hat{a}_\mathrm{r}^\dagger\hat{a}_\mathrm{r} - \chi_\mathrm{qc}\hat{a}_\mathrm{q}^\dagger\hat{a}_\mathrm{q}\hat{a}_\mathrm{c}^\dagger\hat{a}_\mathrm{c} + \frac{\chi'_\mathrm{qc}}{2}\hat{a}_\mathrm{q}^\dagger\hat{a}_\mathrm{q}\hat{a}_\mathrm{c}^{\dagger 2}\hat{a}_\mathrm{c}^{2}
,
\end{eqnarray}
where $\omega_\mathrm{q,c,r}$, $\hat{a}_\mathrm{q,c,r}$, and $K_\mathrm{q,c,r}$ denote the resonance frequencies, annihilation operators, and self-Kerr coefficients of the ancilla qubit, storage cavity, and reservoir resonator, respectively; $\chi_\mathrm{qr}$ and $\chi_\mathrm{qc}$ are the cross-Kerrs between the qubit and the reservoir resonator/storage cavity; and $\chi'_\mathrm{qc}$ is the higher-order cross-Kerr between the ancilla qubit and the storage cavity. Measured values of these parameters are listed in Table~\ref{TableS1}. 

Here, the ancilla transition frequency depends linearly on the cavity photon number via the dispersive shift $\chi_\mathrm{qc}$. This interaction between the ancilla and the storage cavity provides the basis for implementing error correction operations. At the end of each correction cycle, the ancilla is reset through its dispersive coupling ($\chi_\mathrm{qr}$) to the reservoir resonator, which has a large decay rate $\kappa_\mathrm{r}$ and can be used to engineer dissipation for removing the accumulated error entropy.

\begin{table}[t]
\caption{Hamiltonian parameters and coherence times.}
\centering
\begin{tabular}{p{3cm}<{\centering} p{3cm}<{\centering}  p{1.7cm}<{\centering} p{1.7cm}<{\centering} p{1.7cm}<{\centering} p{1.7cm}<{\centering} p{1.7cm}<{\centering} p{1.7cm}<{\centering}}
  \hline
  \hline
    &&&&&&&\\[-0.9em]
  \multicolumn{2}{c}{Measured parameters} &  \multicolumn{2}{c}{Storage cavity ($C$)} & \multicolumn{2}{c}{Transmon qubit ($Q$)} & \multicolumn{2}{c}{Reservoir resonator ($R$)} \\\hline
  &&&&&&&\\[-0.9em]
  \multicolumn{2}{c}{Mode frequency $\omega_\mathrm{c,q,r}/2\pi$ } & \multicolumn{2}{c}{6.611~GHz} & \multicolumn{2}{c}{5.478~GHz} & \multicolumn{2}{c}{8.578~GHz}  \\
  &&&&&&&\\[-0.9em]
  \hline
  &&&&&&&\\[-0.9em]
  \multicolumn{2}{c}{Self-Kerr $K_\mathrm{c,q,r}/2\pi$ } & \multicolumn{2}{c}{1.0~kHz} & \multicolumn{2}{c}{221~MHz} & \multicolumn{2}{c}{-}  \\
  &&&&&&&\\[-0.9em]
  \hline
  &&&&&&&\\[-0.9em]
  \multicolumn{2}{c}{Cross-Kerr $(\chi_\mathrm{qc}, \chi_\mathrm{qr})/2\pi$ } & &\multicolumn{2}{c}{0.88~MHz} & \multicolumn{2}{c}{3.0~MHz} & \\
  &&&&&&&\\[-0.9em]
  \hline  
  &&&&&&&\\[-0.9em]
  \multicolumn{2}{c}{Second-order cross-Kerr $\chi'_\mathrm{qc}/2\pi$ } & & \multicolumn{2}{c}{0.97~kHz} &  \multicolumn{2}{c}{-} & \\
  &&&&&&&\\[-0.9em]
  \hline 
  \hline
  &&&&&&&\\[-0.9em]
  \multicolumn{2}{c}{Relaxation time $T_1$ } & \multicolumn{2}{c}{$1.4~$ms} & \multicolumn{2}{c}{$133~\mu$s} & \multicolumn{2}{c}{113~ns}  \\
  &&&&&&&\\[-0.9em]
  \multicolumn{2}{c}{Pure dephasing time $T_\phi$ } & \multicolumn{2}{c}{$6.5~$ms} & \multicolumn{2}{c}{$388~\mu$s} & \multicolumn{2}{c}{-}  \\
  &&&&&&&\\[-0.9em]
  \multicolumn{2}{c}{Thermal population } & \multicolumn{2}{c}{0.2\%} & \multicolumn{2}{c}{1.2\%} & \multicolumn{2}{c}{-}  \\
  &&&&&&&\\[-0.9em]
  \hline 
  \hline
\end{tabular}
\label{TableS1}
\end{table}

\section{{Unconditional qubit reset method}}

\subsection{Derivation of qubit reset method}

The following derivation extends the quantum bath engineering technique established in previous work~\cite{murch2012} to a system where a qubit is simultaneously coupled to both a storage cavity and a reservoir resonator. After truncating the ancilla Hilbert space to a two-dimensional subspace and neglecting higher-order dispersive interaction, the system Hamiltonian becomes: 
\begin{eqnarray}
    \hat{H} = \omega_{\mathrm{c}} \hat{a}_{\mathrm{c}}^\dagger \hat{a}_{\mathrm{c}}+ \omega_{\mathrm{r}} \hat{a}_{\mathrm{r}}^\dagger \hat{a}_{\mathrm{r}}+ \frac{\omega_{\mathrm{q}}}{2}\hat{\sigma}_z-\frac{\chi_{\mathrm{qc}}}{2}\hat{a}_{\mathrm{c}}^\dagger \hat{a}_{\mathrm{c}}\hat{\sigma}_z - \frac{\chi_{\mathrm{qr}}}{2}\hat{a}_{\mathrm{r}}^\dagger \hat{a}_{\mathrm{r}}\hat{\sigma}_z.
\end{eqnarray}
We apply detuned microwave drives to both the qubit and reservoir resonator with frequencies $\omega_{\mathrm{d}}^{\mathrm{q}}$ and $\omega_{\mathrm{d}}^{\mathrm{r}}$, respectively. Transforming into the rotating frame of these drives yields the Hamiltonian:
\begin{eqnarray}
    \hat{H}_\mathrm{d} &=& \Delta_{\mathrm{r}} \hat{a}^\dagger_{\mathrm{r}} \hat{a}_{\mathrm{r}} + \frac{\Delta_{\mathrm{q}}-\chi_{\mathrm{qc}}\hat{a}^\dagger_{\mathrm{c}} \hat{a}_{\mathrm{c}}}{2}\hat{\sigma}_z -\frac{\chi_{\mathrm{qr}}}{2}\hat{a}^\dagger_{\mathrm{r}} \hat{a}_{\mathrm{r}}\hat{\sigma}_z \notag \\ &+& \frac{\Omega_{\mathrm{R}}}{2}\hat{\sigma}_x + \epsilon_{\mathrm{r}}(\hat{a}^\dagger_{\mathrm{r}}+ \hat{a}_{\mathrm{r}}),
\end{eqnarray}
where $\Delta_{\mathrm{r}} = \omega_{\mathrm{r}}-\omega_{\mathrm{d}}^{\mathrm{r}}$, $\Delta_{\mathrm{q}} = \omega_{\mathrm{q}}-\omega_{\mathrm{d}}^{\mathrm{q}}$, and $\Omega_{\mathrm{R}}$ and $\epsilon_{\mathrm{r}}$ represent the respective drive amplitudes. When considering the resonator decay through the dissipator $\mathcal{D}[\sqrt{\kappa_\mathrm{r}}\hat{a}_\mathrm{r}]$, we arrive at the master equation as $\dot{\rho}=-i[\hat{H}_\mathrm{d},\rho]+\mathcal{D}[\sqrt{\kappa_\mathrm{r}}\hat{a}_\mathrm{r}]$, where $\kappa_\mathrm{r}=1/T_1^\mathrm{r}$ is resonator decay rate and $\mathcal{D}[\hat{L}]\rho=\hat{L}\rho \hat{L}^\dagger-1/2(\hat{L}^\dagger \hat{L}\rho+\rho \hat{L}^\dagger \hat{L})$ is the Lindblad dissipative superoperator with a collapse operator $\hat{L}$. Performing a displaced frame transformation $\hat{D}(\xi_\mathrm{r})=\exp{(\xi_\mathrm{r} \hat{a}_\mathrm{r}^\dagger - \xi_\mathrm{r}^* \hat{a}_\mathrm{r})}$ with $\xi_\mathrm{r}=\frac{-\epsilon_\mathrm{r}}{\Delta_\mathrm{r} -i\kappa_\mathrm{r}/2}$ eliminates all the linear terms of $\hat{a}_{\mathrm{r}}$ in the master equation, giving an effective Hamiltonian, 
\begin{eqnarray}
    \hat{H}_\mathrm{eff} &=& \Delta_{\mathrm{r}} \hat{a}^\dagger_{\mathrm{r}} \hat{a}_{\mathrm{r}} + \frac{\Omega_{\mathrm{R}}}{2}\hat{\sigma}_x+ \frac{\Delta_{\mathrm{q}}-\chi_{\mathrm{qr}}\vert\xi_{\mathrm{r}} \vert^2-\chi_{\mathrm{qc}}\hat{a}^\dagger_{\mathrm{c}} \hat{a}_{\mathrm{c}}}{2}\hat{\sigma}_z -\frac{\chi_{\mathrm{qr}}}{2}(\hat{a}^\dagger_{\mathrm{r}} \hat{a}_{\mathrm{r}}+\xi_{\mathrm{r}} \hat{a}^\dagger_{\mathrm{r}}+\xi_{\mathrm{r}}^* \hat{a}_{\mathrm{r}})\hat{\sigma}_z.
\end{eqnarray}
Transforming into the dressed ancilla basis $\{|\pm\rangle=(|g\rangle\pm|e\rangle)/\sqrt{2}\}$ with a Hadamard rotation and setting $\Delta_{\mathrm{q}} = \chi_{\mathrm{qr}}\vert \xi_{\mathrm{r}}\vert^2 + \delta_{\mathrm{qc}}$ with a frequency shift $\delta_\mathrm{qc}$, we obtain
\begin{eqnarray}
    \hat{H}_\mathrm{eff} = \Delta_{\mathrm{r}} \hat{a}_{\mathrm{r}}^\dagger \hat{a}_{\mathrm{r}} + \frac{\Omega_{\mathrm{R}}}{2}\hat{\sigma}_z- (\frac{\chi_{\mathrm{qc}}}{2}\hat{a}^\dagger_{\mathrm{c}} \hat{a}_{\mathrm{c}}-\delta_{\mathrm{qc}})\hat{\sigma}_x-\frac{\chi_{\mathrm{qr}}}{2}(\hat{a}_{\mathrm{r}}^\dagger \hat{a}_{\mathrm{r}} + \xi_\mathrm{r}\hat{a}^\dagger_{\mathrm{r}}+\xi_\mathrm{r}^* \hat{a}_{\mathrm{r}})\hat{\sigma}_x.
\end{eqnarray}
When $\Delta_{\mathrm{r}}$ is chosen close to the drive amplitudes, the rotating wave approximation allows neglecting the fast-rotating terms involving the qubit and reservoir resonator  ($\hat{a}_{\mathrm{r}}^{\dagger}\hat{a}_{\mathrm{r}}\hat{\sigma}_x$, $\hat{a}_{\mathrm{r}}\hat{\sigma}_-$, $\hat{a}_{\mathrm{r}}^\dagger\hat{\sigma}_+$), yielding:
\begin{eqnarray}
    \hat{H}_\mathrm{eff} = \Delta_{\mathrm{r}} \hat{a}_{\mathrm{r}}^\dagger \hat{a}_{\mathrm{r}} + \frac{\Omega_{\mathrm{R}}}{2}\hat{\sigma}_z- (\frac{\chi_{\mathrm{qc}}}{2}\hat{a}_{\mathrm{c}}^\dagger \hat{a}_{\mathrm{c}} -\delta_{\mathrm{qc}}) \hat{\sigma}_x-\frac{\chi_{\mathrm{qr}}}{2}(\xi \hat{a}_{\mathrm{r}}^\dagger \hat{\sigma}_- + \xi^* \hat{a}_{\mathrm{r}} \hat{\sigma}_+ ).
\end{eqnarray}
To diagonalize the qubit-cavity subsystem, we perform a unitary transformation defined by:
\begin{eqnarray}
\hat{U}&=&\left(\begin{array}{cc}
\cos\hat{\theta} /2 & \sin\hat{\theta} /2 \\
-\sin\hat{\theta} /2 & \cos\hat{\theta} /2 
\end{array}\right),
\end{eqnarray}
with $\tan\hat{\theta}=\chi_{\mathrm{qc}}\left(\hat{a}^{\dagger}_\mathrm{c}\hat{a}_\mathrm{c}-\delta_\mathrm{qc} \right)/\Omega_{\mathrm{R}}$. The diagonalized Hamiltonian is given by:
\begin{eqnarray}
\hat{U}^{\dagger}\hat{H}_\mathrm{eff}\hat{U}=\Delta_{\mathrm{r}}\hat{a}_{\mathrm{r}}^{\dagger}\hat{a}_{\mathrm{r}}+\frac{\hat{\Omega}_{N}}{2}\hat{\sigma}_{z}^{\prime}-\frac{\chi_{\mathrm{qr}}}{4}\xi_{\mathrm{r}}\hat{a}_{\mathrm{r}}^{\dagger}\left(\frac{\Omega_{\mathrm{R}}}{\hat{\Omega}_{N}}+1\right)\hat{\sigma}_{-}^{\prime}-\frac{\chi_{\mathrm{qr}}}{4}\xi_{\mathrm{r}}^{*}\hat{a}_{\mathrm{r}}\left(\frac{\Omega_R}{\hat{\Omega}_{N}}+1\right)\hat{\sigma}_{+}^{\prime},
\end{eqnarray}
where $\hat{\Omega}_{N}=\sqrt{\chi_{\mathrm{qc}}^{2}\left(\hat{a}_{\mathrm{c}}^{\dagger}\hat{a}_{\mathrm{c}}-\delta_\mathrm{qc} \right)^{2}+\Omega_{\mathrm{R}}^{2}}$, $\hat{\sigma}_z^{\prime}$ and $\hat{\sigma}_\pm^{\prime}$ denote Pauli operators in the rotated basis of the qubit. Through employing the adiabatic elimination\cite{reiter2012effective}, we obtain the effective Hamiltonian and collapse operator in the rotated frame:
\begin{eqnarray}
\hat{H^{\prime}}_\mathrm{eff}&=&\frac{\hat{\Omega}_{N}-\Delta_\mathrm{r}}{2}\hat{\sigma}_{z}^{\prime}, \notag \\
\hat{L}_\mathrm{eff}^{\prime}&=&\frac{1}{2}\frac{\chi_\mathrm{qr}|\xi_\mathrm{r}|}{\sqrt{\kappa_\mathrm{r}}}\left(\frac{\Omega_\mathrm{R}}{\widehat{\Omega}_{N}}+1\right)\hat{\sigma}_{-}^{\prime}.\label{Eq:H_eff_rot}
\end{eqnarray}
Finally, transforming the basis back to $\{|+\rangle,|-\rangle\}$ under the assumption $\Omega_\mathrm{R}\gg\chi_{qc}$ ($\hat{\Omega}_\mathrm{N}\approx\Omega_\mathrm{R}$), we obtain
\begin{eqnarray}
\hat{H}_\mathrm{eff}&=&\frac{\Omega_\mathrm{R}-\Delta_\mathrm{r}}{2}\hat{\sigma}_{z}, \notag \\
\hat{L}_\mathrm{eff}&=&\frac{\chi_\mathrm{qr}|\xi_\mathrm{r}|}{\sqrt{\kappa_\mathrm{r}}}\hat{\sigma}_{-}-\frac{\chi_\mathrm{qc}\chi_\mathrm{qr}|\xi_\mathrm{r}|}{2\Omega_\mathrm{R}\sqrt{\kappa_\mathrm{r}}}(\hat{a}_\mathrm{c}^\dagger\hat{a}_\mathrm{c}-\delta_\mathrm{qc})\hat{\sigma}_\mathrm{z}.
\end{eqnarray}
When $\Delta_\mathrm{r}=\Omega_\mathrm{R}$, the dissipative channel drives any ancilla state to $|-\rangle$ at a loss rate $\chi_\mathrm{qr}^2|\xi_\mathrm{r}|^2/\kappa_\mathrm{r}$, as described by the first term in $\hat{L}_\mathrm{eff}$. A subsequent Hadamard operation then rotates $|-\rangle$ to $|g\rangle$, thereby achieving the qubit reset. The additional dephasing term in $\hat{L}_\mathrm{eff}$ is negligible since $\frac{\chi_\mathrm{qc}\chi_\mathrm{qr}|\xi_\mathrm{r}|}{\Omega_\mathrm{R}\sqrt{\kappa_\mathrm{r}}}\sim0$ for sufficiently large drive amplitude $\Omega_\mathrm{R}$. These derivations provide a theoretical foundation for understanding the reset dynamics in our system with dual-resonator coupling.

\begin{figure}[b]
\centering
    \includegraphics{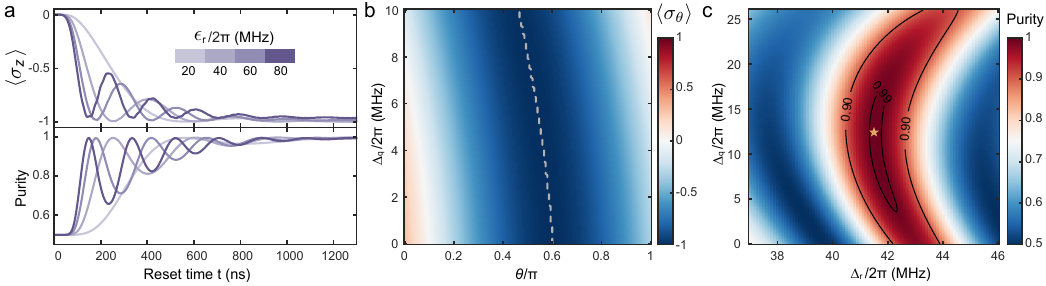}
    \caption{\textbf{Optimization of Reset protocol.} 
    \textbf{a}, Numerical results of the qubit's ground state population (top) and state purity (bottom) as a function of time evolution for various drive amplitudes $\epsilon_\mathrm{r}$.  
    \textbf{b}, Numerically simulated expected value of $\hat{\sigma}_\theta$ as a function of the angle $\theta$ and the qubit drive detuning $\Delta_\mathrm{q}$. The dashed line marks the minimum values for each row.
    \textbf{c}, Numerically simulated purity of the joint cavity-qubit system as a function of $\Delta_\mathrm{q}$ and $\Delta_\mathrm{r}$. The star marks the optimal choice of drive parameters for optimal reset performance in our experimental configuration.
}
    \label{fig:Fig_ResetQ}
\end{figure}

\subsection{Numerical optimization of qubit reset protocol}\label{resetprotocol}
The ancilla reset scheme, illustrated in Fig.2 in the main text, proceeds as follows: First, a Hardmard gate (implemented as a $\pi /2$ pulse) rotates the qubit from the $\hat{\sigma}_z$ basis to the $\hat{\sigma}_x$ basis, mapping the state $\vert g \rangle$ to $\vert -\rangle$ and $\vert e\rangle$ to $\vert + \rangle$. This ensures the reset process minimally affects the ancilla if it is initially in the $\vert g \rangle$ state. Next, simultaneous detuned drives are applied to both the qubit and the reservoir resonator for a specific duration, resetting the qubit to the state $|-\rangle$. Finally, another Hadamard gate returns the qubit back to the $\hat{\sigma}_z$ basis $\{|g\rangle, |e\rangle\}$.

In this section, we perform numerical simulations to validate the reset scheme. To preliminarily verify its effectiveness, we initialize the qubit in the maximally mixed state $\rho_0 =  \frac{1}{2}|g\rangle\langle g| + \frac{1}{2}|e\rangle\langle e| $ and apply the reset drives. With the qubit drive amplitude fixed at $\Omega_\mathrm{R}/2\pi = 40$~MHz, we vary the resonator drive amplitude and finally apply a Hadamard gate to the qubit. As shown in Fig.~\ref{fig:Fig_ResetQ}a, stronger resonator drives shorten the qubit reset time to reach the ground state $|g\rangle$ ($\langle\hat{\sigma}_z\rangle=-1$). The corresponding purity curves provide independent confirmation of this conclusion, offering an alternative to operator-expectation-value monitoring for identifying reset completion in numerical simulations. During the reset process, the resonator drive pulse incorporates rising and falling edges of 190~ns to ensure a sufficiently small photon population in the resonator at the end of the reset operation.

The presence of photons in the storage cavity induces a qubit frequency shift of $\Delta_{\mathrm{q}} = N\chi_\mathrm{qc}$, where $N$ is the number of photons in the storage cavity, resulting in a detuning for the qubit drive pulse. This causes the qubit to stabilize not to a $\hat{\sigma}_z$ eigenstate but to a $\hat{\sigma}_\theta = \cos(\theta) \hat{\sigma}_x + \sin(\theta)\hat{\sigma}_z$ eigenstate. This effect remains negligible when the condition $\Omega_\mathrm{R} \gg \chi_\mathrm{qr}$ is satisfied. We calibrate the relationship between the final stabilized qubit state and the qubit drive detuning for the case of $\Omega_\mathrm{R}/2\pi = 40$~MHz and $\epsilon_\mathrm{r}/2\pi = 80$~MHz, as shown in Fig.~\ref{fig:Fig_ResetQ}b. The results show that for photon numbers $N < 10$, the qubit frequency shift induces only modest changes in $\theta$. In subsequent experiments, quantum optimal control (QOC)~\cite{Khaneja2005} methods are used to compensate for this deviation to ensure the qubit returns to $|g\rangle$ at the end of the reset process.

Finally, we numerically investigate the impact of the reset on the decoherence of quantum states stored in the storage cavity. We first prepare a superposition state $\vert \psi_0\rangle = (\vert 0\rangle + \vert 1\rangle + \vert 2\rangle + \vert 3\rangle +\vert 4\rangle)/\sqrt{5}$ in the storage cavity and then monitor the purity of the joint cavity-qubit state versus $\Delta_{\mathrm{q}}$ and $\Delta_\mathrm{r}$ with fixed $\Omega_\mathrm{R}/2\pi = 40$~MHz and $\epsilon_\mathrm{r}/2\pi = 80$~MHz (see Fig.~\ref{fig:Fig_ResetQ}c). The simulation results show that high coherence can be preserved with optimal parameters. Our experimental configuration uses $\Delta_\mathrm{q}/2\pi = 12.4~\mathrm{MHz}$ and $\Delta_\mathrm{r}/2\pi = 41.5~\mathrm{MHz}$ to achieve optimal reset performance.

\subsection{Calibrating the photon number population of the resonator}
In this section, we analyze the photon number population within the reservoir resonator during the reset process. We first perform calibration measurements to characterize the average photon number $\bar{n}$ inside the reservoir resonator during the reset. This process leads to the resonator achieving a steady-state photon number that induces an a.c. Stark shift in the qubit frequency when an off-resonant drive is applied.

\begin{figure}[tb]
    \includegraphics{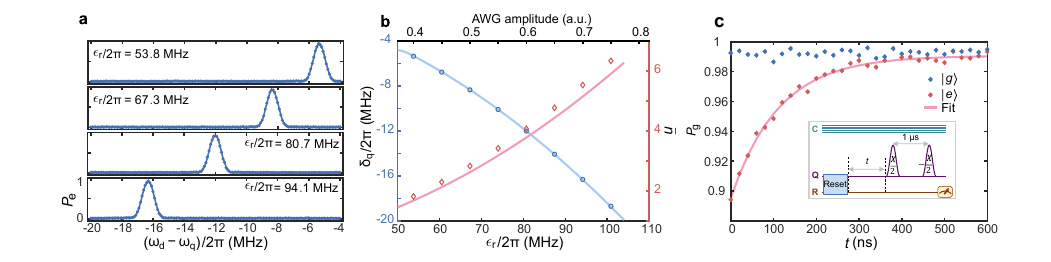}
    \caption{\textbf{Calibration of the reservoir resonator drive.} \textbf{a,} Qubit spectroscopy under different resonator drive amplitudes with fixed drive detuning $\Delta_\mathrm{r}/2\pi = 41.5$ MHz. 
\textbf{b,} Extracted qubit frequency shift as a function of the resonator drive amplitude $\epsilon_\mathrm{r}$. Blue dots represent experimentally measured qubit frequency shifts $\delta_\mathrm{q}$, while the light blue solid line shows simulation results. Red points indicate the average photon number in the resonator estimated using $\bar{n} = \delta_\mathrm{q} / \chi_\mathrm{qr}$, with the light red solid line representing simulated photon numbers. 
\textbf{c,} Measured qubit ground state population from a Ramsey experiment as a function of the waiting time after a reset operation. Blue and red data points correspond to measurements with the qubit initially prepared in the $\vert g\rangle$ and $\vert e\rangle$ states, respectively. The light red solid line shows an exponential fit to the data. The inset illustrates the experimental sequence.
}
    \label{fig:Fig_population_R}
\end{figure}

With fixed $\Delta_\mathrm{r}/2\pi = 41.5$~MHz, we vary the resonator drive amplitude $\epsilon_\mathrm{r}$ and perform qubit spectroscopy to characterize the dependence of the qubit frequency shift $\delta_\mathrm{q}$ on the drive power, as shown in Figs.~\ref{fig:Fig_population_R}a-b. Through comparison with simulations, we establish the correspondence between the arbitrary waveform generator (AWG) output amplitude and the effective drive strength $\epsilon_\mathrm{r}$. The average photon number $\bar{n}$ in the resonator is determined by the relation $\bar{n} = \delta_\mathrm{q} / \chi_\mathrm{qr}$, as shown in Fig.~\ref{fig:Fig_population_R}b. We also present numerical simulation results (solid lines) in the same figure, which show agreement with the experimental results. In our experiment, the drive amplitude is set to $\epsilon_\mathrm{r}/2\pi \approx 80$~MHz, corresponding to an average photon number of $\bar{n} \approx 4$ in the resonator. Then the reset time is approximately 190 ns, as discussed in Sec.~\ref{resetprotocol}.

The residual photon population in the reservoir resonator after the reset can be inferred from a Ramsey experiment performed on the qubit, as illustrated in Fig.~\ref{fig:Fig_population_R}c. By varying the waiting time after the reset, we observe that the resonator contains almost no residual photons if the qubit is reset from the ground state $|g\rangle$. In contrast, a small residual photon population remains if the reset starts from the excited state $|e\rangle$, requiring an additional waiting time of 400~ns for complete dissipation of these photons.

\subsection{Calibration of reset-induced phases}

\begin{figure}[tb]
    \includegraphics{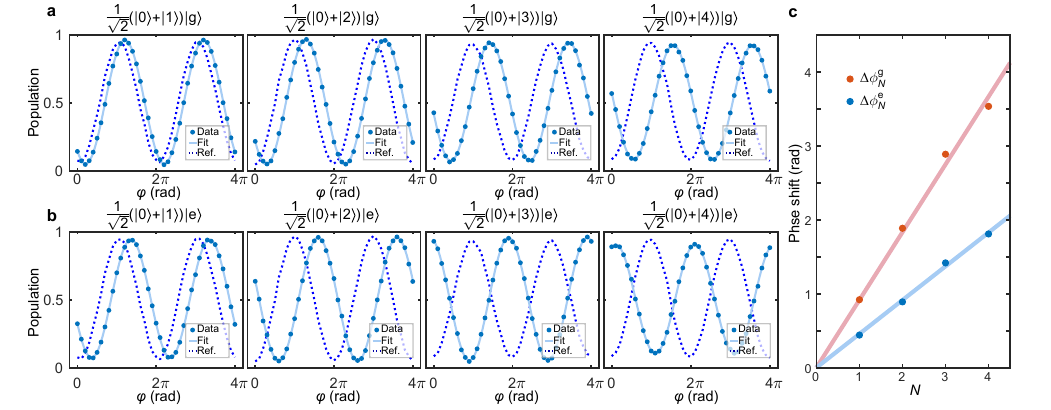}
    \caption{\textbf{Calibration of the reset-induced phase.} \textbf{a-b,} Ramsey interferometry experiments. The cavity and qubit are first prepared in states of the form $\vert 0\rangle + \vert N\rangle$ ($N = 1, 2, 3, 4$) with the qubit in either $\vert g\rangle$ or $\vert e\rangle$. Following the application of the reset operation, the quantum state is decoded from the $\{\vert 0\rangle, \vert N\rangle\}$ subspace of the cavity to the $\{\vert g\rangle, \vert e\rangle\}$ subspace of the qubit to measure the relative phase between $\vert 0\rangle$ and $\vert N\rangle$. The dashed line represents measurements without reset, serving as a baseline. Blue data points show results after a single reset operation, and the solid light-blue line indicates a sinusoidal fit to these data points. \textbf{c,} Extracted phase shifts $\Delta \phi^{g}_N$ and $\Delta \phi^{e}_N$ as a function of cavity photon number $N$. Blue data points represent $\Delta \phi^{g}_N$ with a light-blue linear fit, while red data points show $\Delta \phi^{e}_N$ with a light-red linear fit.
}
    \label{fig:Fig_Reset_phase}
\end{figure}

The reset process induces a deterministic phase shift on the joint cavity-qubit state, which depends on the average photon number in the reservoir resonator and the reset duration. This phase shift, though state-dependent---varying with the qubit state ($|g\rangle$ or $|e\rangle$) and the photon number in the storage cavity---is a constant and can therefore be calibrated and compensated. To quantify this effect, we perform Ramsey interferometry experiments by first initializing the cavity in superposition states $(\vert 0 \rangle + \vert N \rangle)/\sqrt{2}$ for $N=1-4$, with the qubit prepared in either $\vert g \rangle$ or $\vert e \rangle$. After a single reset operation, the cavity state is mapped back onto the qubit through a decoding procedure. Finally, a $\pi/2$ pulse with a variable phase $\varphi$ is applied to the qubit to measure the relative accumulated phase $\Delta \phi^{g ( e)}_N$ between $\vert 0 \rangle$ and $\vert N \rangle$. The experimental results are presented in Figs.~\ref{fig:Fig_Reset_phase}a-b, where the accumulated phases $\Delta \phi^{g}_N$ and $\Delta \phi^{e}_N$ are extracted through sinusoidal fitting to the Ramsey oscillation. As shown in Fig.~\ref{fig:Fig_Reset_phase}c, the relative phases $\Delta \phi^{g (e)}_N$ exhibit a linear dependence on the cavity photon number $N$. Ignoring a negligible decoherence effect on the cavity, the reset operation $\hat{\mathcal{R}}_q$ can be described as:
\begin{eqnarray}\label{Eqn_Reset}
    \hat{\mathcal{R}}_q = \sum _N \left( e^{\Delta\phi_N^{g} }\vert N \rangle\langle N\vert\otimes\vert g\rangle \langle g\vert  +e^{\Delta\phi_N^{e} }\vert N \rangle\langle N\vert\otimes \vert g\rangle \langle e\vert \right),
\end{eqnarray}
on the joint cavity-qubit system. 

To eliminate the influence of reset-induced phase accumulation in AQEC process, we incorporate phase compensation directly into our AQEC procedure by adjusting the phase parameters of the error correction operation preceding the reset. This pre-compensation effectively cancels the accumulated phases during the reset process without adding additional gates to the AQEC protocol.

\subsection{Reset-induced errors on the logical qubit}\label{sec:reset_error}

\begin{figure}[tb]
    \includegraphics{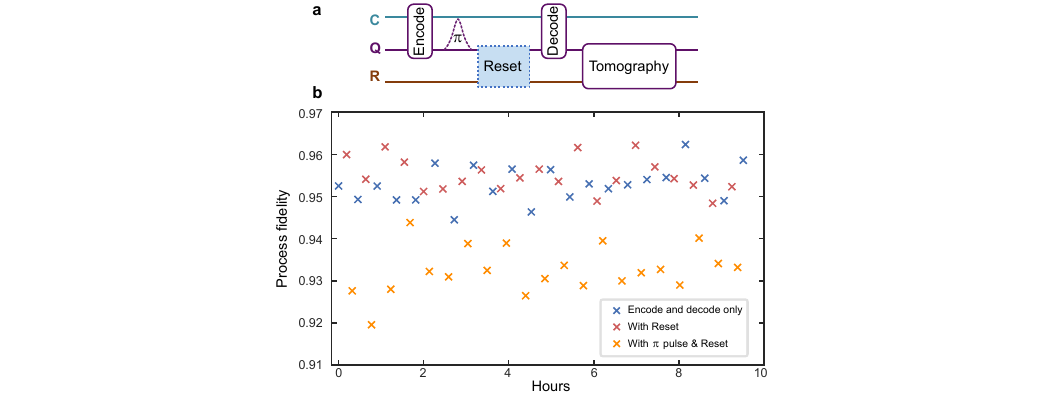}
    \caption{\textbf{Calibration of reset-induced logical qubit errors.} \textbf{a,} Experimental sequence for quantum process tomography used to characterize reset-induced errors. \textbf{b,} Process fidelity over time under three conditions: blue crosses represent direct encoding and decoding without reset; red crosses represent the case with qubit reset; orange crosses represent the case with a qubit flipped from $\vert g \rangle$ to $\vert e \rangle$ followed by reset.} 
    \label{fig:Fig_Reset_protomo}
\end{figure}

As demonstrated in Fig.2 of the main text, the reset operation itself introduces negligible decoherence to the logical qubit state $\vert 0_\mathrm{L} \rangle$. In this section, we further quantify the error induced by integrating the reset operation into an AQEC cycle. For this purpose, we compare the quantum process fidelity across three distinct experimental sequences: (i) the baseline sequence involving only encoding and decoding of the logical state; (ii) a similar sequence with inserting a reset operation between the encoding and decoding processes; and (iii) a similar sequence with inserting an ancilla $\pi$ pulse to flip the qubit state from $\vert g \rangle$ to $\vert e \rangle$ before the reset operation. This last sequence is specifically designed to probe the reset error when the ancilla qubit is in the excited state $\vert e \rangle$. Refer to Fig.~\ref{fig:Fig_Reset_protomo}a for the experimental sequence details.

The results of these measurements, conducted repeatedly over a 10-hour period to ensure statistical reliability, are summarized in Fig.~\ref{fig:Fig_Reset_protomo}b. When the ancilla qubit remains in the ground state, inserting a reset operation yields an average process fidelity of $0.955$, nearly identical to the baseline fidelity of $0.953$. This confirms that the reset process introduces minimal additional error to the logical qubit when the ancilla qubit is in $\vert g \rangle$. In contrast, when the ancilla qubit is flipped to $\vert e \rangle$ prior to the reset, the average process fidelity decreases to $0.933$. This reduction stems from the combined errors of the additional $\pi$ pulse and the reset operation. As a result, these results place an upper bound of $\sim2\%$ for the reset error for the $|e\rangle$ state, implying an average ancilla reset error on the logical qubit below 1\%.

It is important to note that, in all cases, the deterministic phase shifts induced by the reset are pre-calibrated and compensated by correspondingly adjusting the phase of the decoding operation. 

\section{Details of the AQEC procedure}\label{AQEC}
\subsection{AQEC scheme with lowest-order binomial code}\label{AQECbin}
In our experiment, quantum information is encoded in a high-quality superconducting microwave cavity using the lowest-order binomial code~\cite{michael2016}, which provides protection against the dominant error source in this system---the single-photon loss. The logical qubit state is defined as:
\begin{eqnarray}
|\psi_{\mathrm{L}}\rangle
= \alpha\,|0_\mathrm{L}\rangle + \beta\,|1_\mathrm{L}\rangle
= \alpha\,\frac{|0\rangle + |4\rangle}{\sqrt{2}} + \beta\,|2\rangle ,
\end{eqnarray}
with $|\alpha|^2 +|\beta|^2=1$. During idling evolution of duration $T_{\mathrm{W}}$ ($T_\mathrm{W} \ll 1/\kappa_\mathrm{c}$), single-photon loss stochastically transfers the logical states out of the code space into the error space:
\begin{eqnarray}
|\psi_{\mathrm{E}}\rangle = \alpha e^{i\phi_3}|3\rangle + \beta e^{i\phi_1}|1\rangle,
\end{eqnarray}
where $ \phi_1$ and $\phi_3$ represent the accumulated phases on Fock states $|1\rangle$ and $|3\rangle$ due to the self-Kerr nonlinearity over the time interval $T_{\mathrm{W}}$. To correct single-photon loss error, our AQEC operation $\hat{U}_\mathrm{AQEC}$ must both map the erroneous states back to the logical code space and flip the ancilla qubit from $|g\rangle$ to $|e\rangle$. This correction process can be described by the following mapping:
\begin{eqnarray}
\hat{U}_\mathrm{AQEC}|\psi_{\mathrm{E}}\rangle |g\rangle = \left(
\alpha\,\frac{|0\rangle + e^{-i\Delta\phi_{4}^e}|4\rangle}{\sqrt{2}}
+ \beta\,e^{-i\Delta\phi_{2}^e}|2\rangle
\right) |e\rangle,
\end{eqnarray}
where $-\Delta \phi_2^e$ and $-\Delta \phi_4^e$ represent the pre-compensation for the accumulated phases induced by the subsequent qubit reset operation $\hat{\mathcal{R}}_q$, which resets the qubit state from $|e\rangle$ to $|g\rangle$.

Beyond single-photon loss, the binomial logical qubit also experiences continuous deformation due to no-jump evolution, governed by the operator $e^{-(\kappa_\mathrm{c}/2) \hat{a}_\mathrm{c}^\dagger \hat{a}_\mathrm{c} t}$. Here, $\kappa_\mathrm{c}$ represents the intrinsic photon decay rate of the cavity mode. For sufficiently short evolution time ($\kappa_\mathrm{c}T_\mathrm{W}\ll 1$), the final state can be approximated as:
\begin{eqnarray}
|\psi'_{\mathrm{L}}\rangle
= \alpha\left(\cos\theta|0\rangle + \sin\theta e^{i\phi_{4}}|4\rangle\right)
+ \beta e^{i\phi_2}|2\rangle ,
\end{eqnarray}
with $\theta$ defined by $\tan\theta = e^{-2\kappa_\mathrm{c} T_{\mathrm{W}}}$. Even when no photon loss is detected, this process asymmetrically reduces the occupation of higher Fock states, leading to a distortion of the encoded quantum information. Fortunately, this no-jump distortion can also be reversed by the same AQEC operation:
\begin{eqnarray}
\hat{U}_\mathrm{AQEC}|\psi'_{\mathrm{L}}\rangle|g\rangle
= \left(
\alpha\,\frac{|0\rangle + e^{-i\Delta\phi_{4}^g}|4\rangle}{\sqrt{2}}
+ \beta\,e^{-i\Delta\phi_{2}^g}|2\rangle
\right) |g\rangle,
\end{eqnarray}
with $-\Delta \phi_2^g$ and $-\Delta \phi_4^g$ representing the pre-compensated phases for reset operations initiating from the $|g\rangle$ state. Therefore, the AQEC control pulse is subject to two principal constraints designed to counteract both jump-induced error and the no-jump evolution effect. After applying the AQEC operation $\hat{U}_{\mathrm{AQEC}}$ and the reset operation $\hat{\mathcal{R}}_q$ from Eq.~\ref{Eqn_Reset}, the outcome of a complete error correction can be expressed as~\cite{ma2020}:
\begin{eqnarray}
|\psi'_{\mathrm{L}}\rangle|g\rangle \oplus |\psi_{\mathrm{E}}\rangle|g\rangle
\xrightarrow{\hat{U}_{\mathrm{AQEC}}} |\psi_{\mathrm{L}}^g\rangle|g\rangle \oplus|\psi_{\mathrm{L}}^e\rangle|e\rangle  \xRightarrow{\hat{\mathcal{R}}_q} |\psi_{\mathrm{L}}\rangle|g\rangle,
\end{eqnarray}
where $|\psi_{\mathrm{L}}^g\rangle$ and $|\psi_{\mathrm{L}}^e\rangle$ represent the logical qubit states with pre-compensated reset phases for ancilla states $|g\rangle$ and $|e\rangle$, respectively. This shows that whether the system experienced no-jump evolution ($|\psi'_{\mathrm{L}}\rangle$) or a single-photon loss error ($|\psi_{\mathrm{E}}\rangle$), the logical qubit is ultimately restored to its correct logical state after a complete error correction. In the next section, we will discuss how the unitary operation $\hat{U}_{\mathrm{AQEC}}$ is implemented using quantum optimal control techniques.

\subsection{Quantum optimal control}\label{sec:grape}

\begin{figure}[b]
    \includegraphics{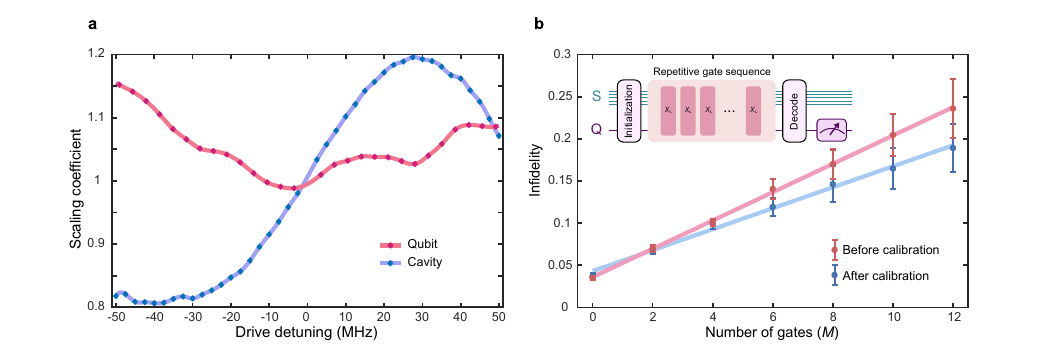}
    \caption{\textbf{Calibration of quantum optimal control pulse.} \textbf{a,} Calibration of the scaling coefficient of the pulse amplitude for both qubit and cavity drives. The data points represent the measured calibration coefficients, while the smooth curve was generated using cubic spline interpolation, ensuring continuity of the function and its first two derivatives across all data points. \textbf{b,} Average infidelity as a function of the number of gates $M$. Data points with error bars represent results before (red) and after (blue) calibration, while solid lines indicate corresponding linear fits. The inset shows the pulse sequence used in this experiment.
}
    \label{fig:Fig_GRAPE}
\end{figure}

The control pulses for all quantum operations in our experiment---including the encoding and decoding of the bosonic logical qubit, AQEC operations, and optimal qubit pulses for frame transformations---are designed using the gradient ascent pulse engineering (GRAPE) algorithm~\cite{Khaneja2005}. This approach maximizes the state-transfer fidelity defined as $F = \left| \sum_i \langle \psi_i^f | \hat{U}_{\text{ideal}} | \psi_i^0 \rangle \right|^2$, where $\hat{U}_{\text{ideal}}$ represents the target unitary operation that simultaneously maps each initial state $|\psi_i^0\rangle$ to its corresponding final state $|\psi_i^f\rangle$. Optimized pulses are applied simultaneously to both the qubit and the cavity mode to achieve high-fidelity and fast control.

To implement these optimized pulses, we consider two distinct approaches. The first approach leverages piecewise constant pulses, where each pulse is digitized with a time resolution of $\Delta t$, and the amplitude at each time step is treated as an optimizable parameter. However, this time-domain parameterization faces significant experimental challenges: the generated AWG waveforms are subject to distortion during propagation through the electronic circuitry, and the complex and irregular temporal shapes of GRAPE-optimized pulses make experimental calibration particularly difficult. To overcome these limitations, we adopt an alternative strategy that employs frequency-domain parameterization as optimization variables~\cite{Machnes2018}. This approach utilizes a basis set of band-limited functions, enabling the generation of control sequences that are more robust and amenable to calibration in practical experimental settings. Specifically, the qubit and cavity control pulses $\epsilon_{\mathrm{q}}^{I(Q)}(t)$ and $\epsilon_{\mathrm{c}}^{I(Q)}(t)$ are parameterized in the frequency domain using Fourier basis decomposition, which can be expressed as:
\begin{eqnarray}
\epsilon_{\mathrm{q}}^{I}(t) &=& \left\{A_0^{I,\mathrm{q}} +\sum_{k=1}^{M} \left[ A_k^{I,\mathrm{q}} \cos(2\pi f_k t) + B_k^{I,\mathrm{q}} \sin(2\pi f_k t) \right]\right\} \hat{\sigma}_x, \notag \\ \epsilon_{\mathrm{q}}^{Q}(t) &=& \left\{A_0^{Q,\mathrm{q}} +\sum_{k=1}^{M} \left[ A_k^{Q,\mathrm{q}} \cos(2\pi f_k t) + B_k^{Q,\mathrm{q}} \sin(2\pi f_k t) \right]\right\} \hat{\sigma}_y,
\end{eqnarray}
and 
\begin{eqnarray}
\varepsilon_{\mathrm{c}}^{I}(t) &=& \left\{A_0^{I,\mathrm{c}} +\sum_{k=1}^{M} \left[ A_k^{I,\mathrm{c}} \cos(2\pi f_k t) + B_k^{I,\mathrm{c}} \sin(2\pi f_k t) \right]\right\} (\hat{a}_\mathrm{c}^\dagger+\hat{a}_\mathrm{c}), \notag \\ \varepsilon_{\mathrm{c}}^{Q}(t) &=& \left\{A_0^{Q,\mathrm{c}} +\sum_{k=1}^{M} \left[ A_k^{Q,\mathrm{c}} \cos(2\pi f_k t) + B_k^{Q,\mathrm{c}} \sin(2\pi f_k t) \right] \right\} (i\hat{a}_\mathrm{c}^\dagger-i\hat{a}_\mathrm{c}),
\end{eqnarray}
where $f_k = k\Delta f$ represents the frequency components with $\Delta f$ being the frequency spacing. The coefficients $A_k^{I(Q),\mathrm{q(c)}}$ and $B_k^{I(Q),\mathrm{q(c)}}$ represent the amplitudes of each frequency component and serve as the optimizable parameters in our method.

This parameterization enables a practical calibration approach for the driving pulses. We calibrate driving strengths at a set of sparse sample frequencies using the same method developed in Ref.~\cite{ni2023}. This method employs the Ramsey oscillation experiment on the qubit to measure the a.c. Stark shifts induced by drives at these specific detunings. A smooth functional representation of the sparse calibration data is then obtained through cubic spline interpolation, which allows us to construct a continuous correction function across the entire frequency band of interest for experimental waveform calibration. The calibration results of the correction coefficients for both qubit and cavity drives across frequency detunings from -50 MHz to 50 MHz are shown in Fig.~\ref{fig:Fig_GRAPE}a.

To evaluate the performance, we implement a benchmarking experiment by applying $M$ logical $X_\mathrm{L}$ gates to the binomial code state $\left\vert 0_\mathrm{L}\right\rangle$. Finally, we perform decoding and characterize the output state by evaluating its fidelity with respect to the ideal state. As shown in Fig.~\ref{fig:Fig_GRAPE}b, the average infidelity exhibits linear growth with increasing $M$, enabling extraction of the single-operation infidelity through linear fitting. The calibrated pulses achieve an average infidelity of $1.24 \pm 0.11\%$, representing a $26\%$ reduction over the uncalibrated baseline of $1.68 \pm 0.05\%$. These results demonstrate the effectiveness of our frequency-domain calibration approach for achieving high-fidelity quantum optimal control.

\begin{figure}[b]
    \includegraphics{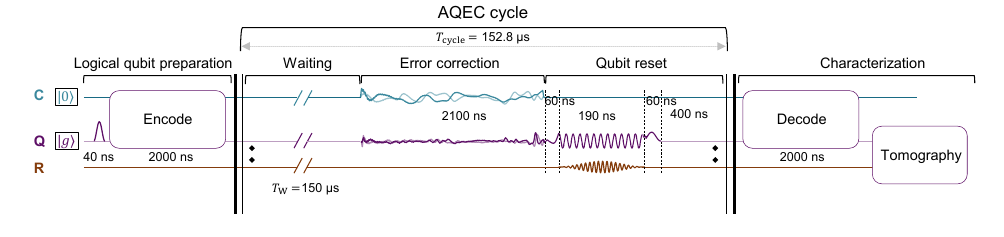}
    \caption{\textbf{Experimental sequence for the AQEC process.}} 
    \label{fig:Figure_AQEC_sequence}
\end{figure}

\subsection{Experimental protocol for the repetitive AQEC}
The repetitive AQEC protocol begins with the initialization of both the ancilla qubit and the storage cavity to their ground states. The ancilla is prepared in the $|g\rangle$ state through post-selection, while the cavity is initialized to the vacuum state $|0\rangle$ through a parity measurement and post-selection. This initialization procedure effectively eliminates residual thermal excitations present in both quantum systems prior to the experimental sequence.

The quantum process tomography of the repetitive AQEC cycle is implemented through the following procedure. First, the ancilla qubit is prepared in four distinct input states \{$|g\rangle$, $|e\rangle$, $(|g\rangle +|e\rangle)/\sqrt{2}$, $(|g\rangle - i|e\rangle)/\sqrt{2}$\}. Each qubit state is then encoded into the binomial logical states \{ $\ket{0_\mathrm{L}}$, $\ket{1_\mathrm{L}}$, $\ket{+X_\mathrm{L}}$, $\ket{-Y_\mathrm{L}}$\} of the storage cavity using a numerically optimized pulse of approximately 2000 ns duration. The encoded logical information is subsequently protected through multiple AQEC cycles before being decoded back to the ancilla for characterization to reconstruct the complete quantum process. The experimental sequence of the full AQEC process is depicted in Fig.~\ref{fig:Figure_AQEC_sequence}. A single AQEC cycle has a total duration of approximately $152.8~\mathrm{\mu s}$, composed of three main components: (i) an idling duration of $T_{\mathrm{W}} = 150~\mathrm{\mu}$s, optimized to balance operational errors, no-parity-jump backaction errors, and photon-loss errors; (ii) error correction operations implemented through a 2100~ns GRAPE pulse; and (iii) a 710~ns qubit reset process to dissipate the error entropy in the ancilla qubit.

\begin{figure}[t]
    \includegraphics{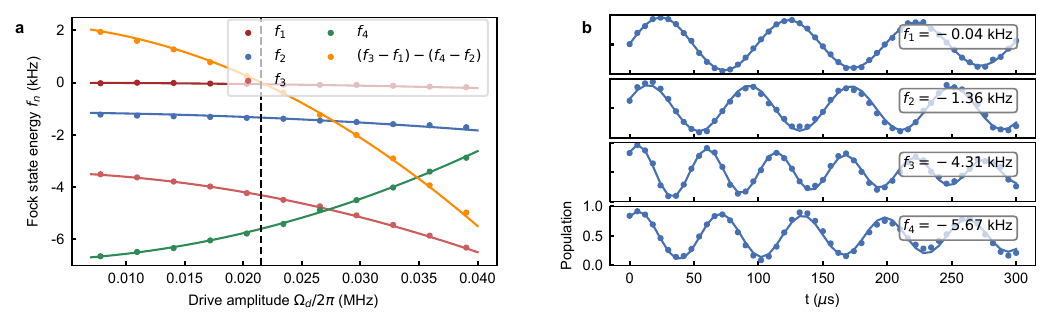}
    \caption{\textbf{Eliminating the dephasing effect with the PASS drive.} 
\textbf{a,} Measured phase accumulation rates $f_n$ for Fock states $\ket{n}$ ($n=1,2,3,4$) relative to $\ket{0}$, as well as their difference $\Delta f = (f_3 - f_1) - (f_4 - f_2)$. 
\textbf{b,} Measured Ramsey oscillations between Fock states $\ket{n}$ ($n=1,2,3,4$) and $\ket{0}$ under optimal PASS drive amplitude.
}
    \label{fig:Fig_PASS}
\end{figure}

To actively suppress logical codeword dephasing induced by self-Kerr nonlinearity, we employ the PASS technique developed in Ref.~\cite{ma2020} during the waiting interval of each AQEC cycle. In this method, we apply an off-resonant drive pulse with 1000-ns rising and falling edges to the ancilla qubit, creating photon-number-dependent frequency shifts that counteract the intrinsic self-Kerr effect. We characterize the PASS technique by measuring relative phase accumulation rates $f_n$ for Fock states $\ket{n}$ ($n = 1, 2, 3, 4$) referenced to the vacuum state \(\ket{0}\), with the drive frequency detuned by $-3.5\chi_\mathrm{qc}$. As shown in Fig.~\ref{fig:Fig_PASS}a, the measured rates as a function of the drive amplitude reveal an optimal operating point at $\Omega_\mathrm{d}/2\pi = 0.0215~\mathrm{MHz}$, where the extracted rates (see Fig.~\ref{fig:Fig_PASS}b) satisfy the error-transparent condition:
\begin{eqnarray}
(f_4 - f_2) - (f_3 - f_1) \approx 0,
\end{eqnarray}
thereby effectively canceling Kerr-induced dephasing of the logical qubit.

The experimental protocol presented here provides a general framework for repetitive AQEC operations. These methodologies constitute a code-agnostic toolkit that can be systematically applied to diverse bosonic codes through individual parameter optimization for each code. In the following section, we demonstrate this versatility by implementing our AQEC scheme on alternative logical encodings, confirming the broad applicability of the method.

\subsection{Applicability of AQEC scheme to diverse bosonic codes}\label{sec:AQEC_other}

Our AQEC scheme employs a single unitary operation to simultaneously recover the logical qubit from the error space back to the code space and transfer the error entropy to the ancilla, where it is removed via engineered dissipation. This method, which does not require explicit syndrome measurements, significantly broadens its applicability to diverse bosonic codes, including those lacking simple stabilizer structures. We demonstrate this generalizability through experimental implementations of two distinct bosonic codes: numerically optimized $\sqrt{17}$ bosonic code~\cite{michael2016} and an approximate QEC code with two-component Fock-state encodings~\cite{cai2021PRL,wang2022NC}.

\begin{figure*}[b]
    \includegraphics{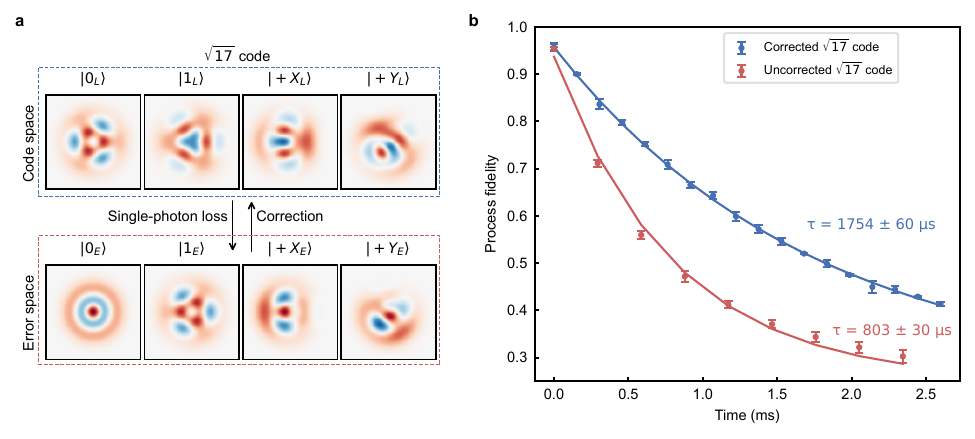}
    \caption{\textbf{Performance of repetitive AQEC operations for the $\sqrt{17}$ code.} 
\textbf{a}, Wigner function representations of the $\sqrt{17}$ code logical states: $|0_\mathrm{L}\rangle = (\sqrt{7-\sqrt{17}}|0\rangle + \sqrt{\sqrt{17}-1}|3\rangle)/\sqrt{6}$ and $|1_\mathrm{L}\rangle = (\sqrt{9-\sqrt{17}}|1\rangle - \sqrt{\sqrt{17}-3}|4\rangle)/\sqrt{6}$. The corresponding error states for single-photon loss are $|0_{\mathrm{E}}\rangle = |2\rangle$ and $|1_{\mathrm{E}}\rangle = (\sqrt{\sqrt{17}-1}|0\rangle - \sqrt{7-\sqrt{17}}|3\rangle)/\sqrt{6}$. \textbf{b}, Process fidelity as a function of the storage time of the $\sqrt{17}$ code with and without AQEC protection. Error bars represent one standard deviation obtained from repeated measurements. All curves are fitted to the function $F_{\chi} = A e^{-t/\tau} + 0.25$ to extract the characteristic lifetimes $\tau$ with uncertainties derived from the fitting procedure.}
    \label{fig:Fig_Sqrt17_QECresults}
\end{figure*}

We first examine the $\sqrt{17}$ code (see Fig.~\ref{fig:Fig_Sqrt17_QECresults}a), a numerically optimized bosonic code that achieves the minimum average photon number among bosonic codes capable of exact correction of single-photon loss error~\cite{michael2016}. The $\sqrt{17}$ codewords are defined as:
\begin{eqnarray}
|0_\mathrm{L}\rangle &=& (\sqrt{7-\sqrt{17}}|0\rangle + \sqrt{\sqrt{17}-1}|3\rangle)/\sqrt{6}, \notag \\ |1_\mathrm{L}\rangle &=& (\sqrt{9-\sqrt{17}}|1\rangle - \sqrt{\sqrt{17}-3}|4\rangle)/\sqrt{6}.
\end{eqnarray}
Once a single-photon loss occurs, the logical states will be transferred into the error space spanned by
\begin{eqnarray}
|0_\mathrm{E}\rangle &=& |2\rangle, \notag \\ |1_\mathrm{E}\rangle &=& (\sqrt{\sqrt{17}-1}|0\rangle - \sqrt{7-\sqrt{17}}|3\rangle)/\sqrt{6}.
\end{eqnarray}
Compared to the lowest-order binomial code with an average photon number of 2, the $\sqrt{17}$ code exhibits a smaller average photon number of $1.56$, and thus simultaneously reduces the photon-jump error rate and the error rate induced by no-jump evolution. The absence of a well-defined parity structure in the $\sqrt{17}$ codewords renders conventional parity-based syndrome detection inapplicable. This limitation, however, is directly addressed by our measurement-free correction scheme. 

We implement a similar AQEC procedure for the $\sqrt{17}$ code with the binomial code in section~\ref{AQECbin}. During the idling time in the AQEC, we apply a different PASS drive with a detuning $-1.325\chi_\mathrm{qc}$ and an amplitude $\Omega_\mathrm{d}/2\pi = 0.039~\mathrm{MHz}$ to satisfy two error-transparent conditions for suppressing dephasing errors. The first condition, $(f_3-f_2)-f_1=0$, ensures active suppression of self-Kerr-induced dephasing, mirroring the protection mechanism employed for the binomial code. The second condition, $(f_4-f_1)-f_3=0$, addresses a critical challenge specific to the $\sqrt{17}$ code: the self-Kerr nonlinearity incrementally increases the overlap between the logical state $\vert0_{\mathrm{L}}\rangle$ and the error state $\vert1_{\mathrm{E}}\rangle$, which would otherwise severely undermine error-correction capability. 

\begin{figure*}[t]
    \includegraphics{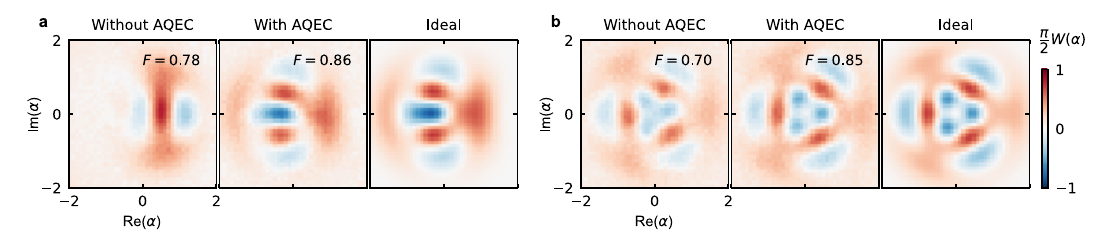}
    \caption{\textbf{Protection of logical states by AQEC.} \textbf{a}, Measured Wigner functions for the $\sqrt{17}$ code without (left) and with (middle) AQEC protection, as well as the ideal $|+X_\mathrm{L}\rangle$ state (right). \textbf{b}, Measured Wigner functions for the $\{|1\rangle, |4\rangle\}$ code without (left) and with (middle) AQEC protection, as well as the ideal $|+X_\mathrm{L}\rangle$ state (right).}
    \label{fig:Fig_othercodewigner}
\end{figure*}

Under repetitive AQEC, the $\sqrt{17}$ code achieves a 2.2-fold lifetime extension relative to the uncorrected logical qubit and surpasses the break-even point by 2\% (Fig.~\ref{fig:Fig_Sqrt17_QECresults}b). Additionally, the measured Wigner functions of the logical state $|+X_\mathrm{L}\rangle$ with and without AQEC protection are presented in Fig.~\ref{fig:Fig_othercodewigner}a, confirming effective quantum information preservation under AQEC protection.

Secondly, we demonstrate AQEC implementation on the approximate QEC code spanned by $\{|1\rangle, |4\rangle\}$, specifically designed for quantum metrology applications. This encoding leverages the large photon number separation between logical states to enhance susceptibility to external parameter variations, making it particularly suitable for detecting frequency perturbations and related sensing tasks. Our AQEC scheme can be used to correct single-photon loss errors during the sensing process using this approximate QEC code. The measured Wigner functions of the Fock superposition state $(|1\rangle+|4\rangle)/\sqrt{2}$ with and without AQEC protection are presented in Fig.~\ref{fig:Fig_othercodewigner}b. Compared to the unprotected case, the phase coherence of the Fock states is significantly improved, demonstrating the effective coherence preservation, which is critical for achieving high-precision sensing.

\subsection{Error analysis}
To investigate the limiting factors in our AQEC experiments, we employ numerical simulations based on solving the master equation to comprehensively analyze and understand the measured results. Here, the reduction in the normalized process fidelity, defined as $\varepsilon=1-(F_\mathrm{\chi}-0.25)/0.75$, is used to quantify the impact of various error sources on a single AQEC cycle, as summarized in Table~\ref{Table_error}. As introduced in Sec.~\ref{AQEC}, a single AQEC cycle consists of three stages: the idling evolution, the GRAPE-based error correction operation, and the final ancilla reset process for entropy removal. Accordingly, the error budget in the AQEC experiment includes encoding-decoding errors, intrinsic errors, gate errors, and reset errors. Below, we analyze each error source individually for the AQEC experiments based on both the lowest-order binomial code and $\sqrt{17}$ code.

Encoding-decoding errors primarily originate from inaccuracies in system initialization, the information transfer processes assisted by ancillary qubits, and the final ancilla readout. While these errors do not affect the logical qubit lifetime, they introduce systematic bias in the process fidelity. 

Intrinsic errors---those beyond the correctable error set $\{\hat{I},\hat{a}\}$ for both the lowest-order binomial and $\sqrt{17}$ codes---include multi-photon losses, single photon gain due to environmental thermal excitation, and dephasing. Since both photon loss and gain rates are proportional to the average photon number in the codeword space, $\sqrt{17}$ code (with an average photon number $\bar{n}_\mathrm{c}\approx1.56$) consequently exhibits lower error rates from these sources compared to the lowest-order binomial code (with an average photon number $\bar{n}_\mathrm{c}=2$), as shown in Table~\ref{Table_error}. 

\begin{figure*}[tb]
    \centering
    \includegraphics{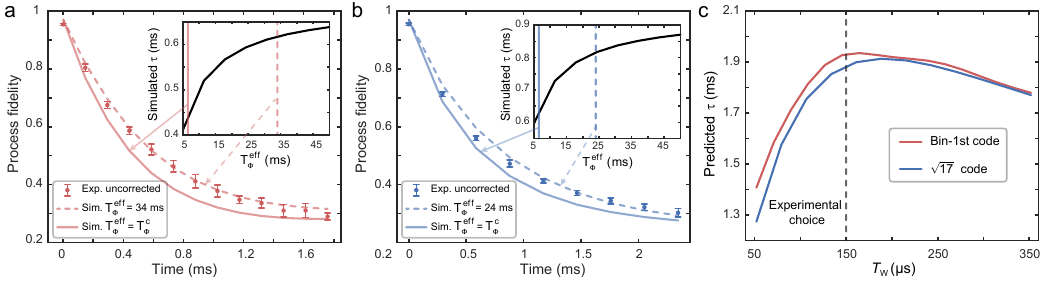}
    \caption{\textbf{Estimations of the effective pure dephasing times and predictions of the logical qubit lifetime.}
    \textbf{a-b}, Process fidelity as a function of time under different effective dephasing times for the lowest-order binomial code (a) and $\sqrt{17}$ code (b). Inset: simulated logical lifetime as a function of the effective pure dephasing time.
    \textbf{c}, The numerically predicted logical lifetimes for the lowest-order binomial and $\sqrt{17}$ codes as a function of the idling time between AQEC cycles. The extracted effective dephasing times from (a-b) are used in the simulation.
    }
    \label{fig:AQEC_tw}
\end{figure*}

Dephasing errors in the AQEC experiment can be categorized into two distinct types: intrinsic pure dephasing of the cavity itself, and externally-induced dephasing, including ancilla-excitation-induced dephasing and Kerr-induced dephasing. By comparing numerical simulation results with experimental data, we note that the effective pure dephasing time in the AQEC differs from that measured using the Fock superposition state $(|0\rangle+|1\rangle)/\sqrt{2}$. We therefore adjust the pure dephasing time in the AQEC simulations to match the measured process fidelity decay curves of the uncorrected logical qubits. We extract the effective pure dephasing times $T^{\mathrm{eff}}_\phi$ of 34~ms and 24~ms for the binomial and $\sqrt{17}$ codes, respectively (Figs.~\ref{fig:AQEC_tw}a-b), corresponding to infidelity contributions of 2.03\% and 2.10\%. Externally-induced dephasing mainly stems from two sources: ancilla thermal excitation and residual Kerr effect. The error from ancilla-induced dephasing is quantified by the formula $n_{\mathrm{th}}^\mathrm{q}(1-e^{-T_\mathrm{cycle}/T_1^\mathrm{q}})$, where $n_{\mathrm{th}}^\mathrm{q}$ is the ancilla thermal population. Both codes exhibit comparable contributions from this mechanism due to similar cycle time $T_\mathrm{cycle}$. Although Kerr-induced dephasing is, in principle, eliminated using the PASS method, experimental imperfections inevitably yield residual Kerr-induced dephasing in practice. Moreover, the PASS drive tends to induce additional qubit excitations, which similarly lead to dephasing. From numerical simulations, we quantify these combined errors of 0.1\% and 1.1\% for the lowest-order binomial code and $\sqrt{17}$ code, respectively. In the Table~\ref{Table_error}, we represent the total dephasing errors from these two error sources. 

The AQEC gate errors are evaluated by comparing numerical simulations (incorporating decoherence and experimental waveforms) with the ideal AQEC operator $\hat{U}_{\mathrm{AQEC}}$, as presented in Table~\ref{Table_error}.

Reset errors, as established in Sec.~\ref{sec:reset_error}, are negligible on the logical state when the ancilla qubit remains in the ground state, but induce a 2\% process fidelity loss on the logical qubit when the ancilla is in the excited state. For a single AQEC cycle of the lowest-order binomial code, the reset error is calculated as $2\%*(1-e^{-T_\mathrm{W}\bar{n}_\mathrm{c}/T_1^\mathrm{c}})$. The same analytical methodology can be directly applied to $\sqrt{17}$ code, yielding errors that remain within a comparable range.

Based on the above error analysis, we numerically investigate the relationship between the theoretically predicted logical qubit lifetime $\tau$~\cite{ni2023} and the waiting time $T_\mathrm{W}$ for both the lowest-order binomial and $\sqrt{17}$ codes. As shown in Fig.~\ref{fig:AQEC_tw}c, our experimentally chosen operating point enables optimal AQEC performance for this system.

\begin{table}[b]
\caption{Error budget for AQEC processes.}
\begin{tabular}{p{2cm}<{\centering} p{3cm}<{\centering} p{2cm}<{\centering} p{2.5cm}<{\centering} p{2cm}<{\centering} p{2cm}<{\centering} p{2cm}<{\centering}}
  \hline
  \hline
  Error source & Multi-photon losses and gain & 
  Intrinsic dephasing & Induced dephasing & Gate error & Reset error & Total error \\\hline
  &\\[-0.9em]
  Bin. code  & 2.47\% & 2.03\% & 0.92\% & $\sim$1.82\% & $\sim$0.39\% & $\sim$7.63\% \\
  &\\[-0.9em]
  $\sqrt{17}$ code & 1.97\% & 2.10\% & 1.92\% & $\sim$1.79\% & $\sim$0.31\% & $\sim$8.09\%\\
  &\\[-0.9em]
  \hline
  \hline
\end{tabular}
\label{Table_error}
\end{table}

\subsection{AQEC scheme for correcting combined types of errors}
In this subsection, we discuss the extension of our AQEC scheme to protect the logical qubit against multiple error types simultaneously. While previous implementations focused solely on single-photon loss errors, practical implementations for achieving better QEC performance require handling multiple types of errors, such as multi-photon losses and dephasing errors. For instance, correcting both single- and two-photon loss errors demands logical codewords capable of distinguishing these error types, such as utilizing the higher-order binomial code~\cite{michael2016}:
\begin{eqnarray}
\left\vert 0_\mathrm{L}\right\rangle &=& \frac{1}{2}(\left\vert 0\right\rangle + \sqrt{3}\left\vert 6\right\rangle), \notag \\
\left\vert 1_\mathrm{L}\right\rangle &=& \frac{1}{2}(\sqrt{3}\left\vert 3\right\rangle + \left\vert 9\right\rangle).
\end{eqnarray}
When the logical qubit experiences single-photon loss ($\hat{a}_\mathrm{c}$) or two-photon loss ($\hat{a}^2_\mathrm{c}$), the logical basis states $\left\vert 0_\mathrm{L}\right\rangle$ and $\left\vert 1_\mathrm{L}\right\rangle$ are mapped to distinct error subspaces $\mathcal{E}_1^{\downarrow}$ spanned by $\{\left\vert 0_\mathrm{E_1}\right\rangle = \left\vert 5\right\rangle,\left\vert 1_\mathrm{E_1}\right\rangle = \frac{1}{\sqrt{2}}(\left\vert 2\right\rangle+\left\vert 8\right\rangle)\}$ for single-photon loss and $\mathcal{E}_2^\downarrow$ spanned by $\{\left\vert 0_\mathrm{E_2}\right\rangle = \left\vert 4\right\rangle,\left\vert 1_\mathrm{E_2}\right\rangle = \frac{1}{\sqrt{5}}(\left\vert 1\right\rangle+2\left\vert 7\right\rangle)\}$ for two-photon loss. 

\begin{figure*}[tb]
    \includegraphics{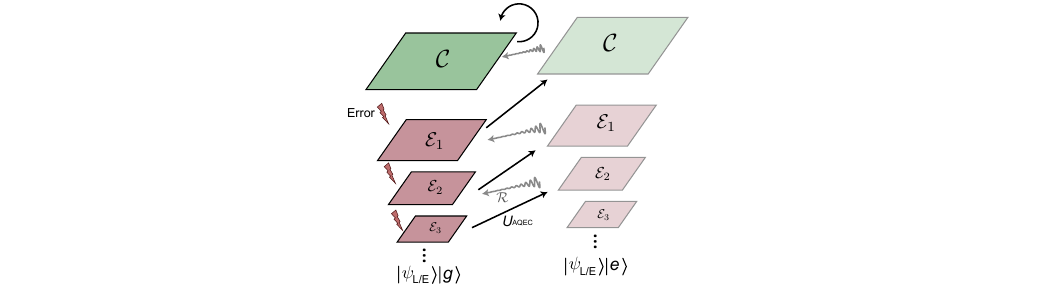}
    \caption{\textbf{Schematic of the cascaded AQEC protocol}.  Higher-order logical qubit errors are sequentially corrected through intermediate error subspaces before final recovery to the logical code space.}
    \label{fig:Fig_cAQEC}
\end{figure*}

In order to simultaneously correct these errors, we here develop a cascaded AQEC scheme, as illustrated in Fig.~\ref{fig:Fig_cAQEC}. In this scheme, we first correct the two-photon loss error by mapping error states from the $\mathcal{E}_2^\downarrow$ subspace back to the single-photon-loss error subspace $\mathcal{E}_1^\downarrow$, then correct single-photon loss errors by mapping error states from $\mathcal{E}_1^\downarrow$ back to the logical subspace $\mathcal{C}$. The complete AQEC unitary operation simultaneously achieves the state transfers:
\begin{eqnarray}\label{eq_cAQEC}
\hat{U}_\mathrm{AQEC}|\psi_\mathrm{E_2}\rangle|g\rangle &=& |\psi_\mathrm{E_1}\rangle|e\rangle, \notag \\
\hat{U}_\mathrm{AQEC}|\psi_\mathrm{E_1}\rangle|g\rangle &=& |\psi_\mathrm{L}\rangle|e\rangle, \notag \\ 
\hat{U}_\mathrm{AQEC}|\psi'_\mathrm{L}\rangle|g\rangle &=& |\psi_\mathrm{L}\rangle|g\rangle, 
\end{eqnarray}
where $|\psi_\mathrm{E_2}\rangle$, $|\psi_\mathrm{E_1}\rangle$, and $|\psi'_\mathrm{L}\rangle$ represent the logical qubit states after the occurrence of two-photon loss errors, single-photon loss errors, and no-jump non-unitary evolution, respectively. Following each correction step, a reset operation returns all ancilla qubit states to $|g\rangle$, completing the AQEC cycle. Phase compensation in $\hat{U}_\mathrm{AQEC}$ for reset-induced phase shifts, though omitted here for clarity, is incorporated in practical implementations.

This cascaded AQEC scheme implements a stepwise recovery process where errors are corrected sequentially through a hierarchy of error subspaces~\cite{shtanko2025}, thereby minimizing the operational overhead. While not achieving complete error correction in a single QEC cycle, this approach significantly reduces the number of required operations compared to schemes attempting simultaneous correction of all types of errors. 

This modular framework can readily extend to handle combined types of errors, including photon loss ($\hat{a}_\mathrm{c}$), photon gain ($\hat{a}_\mathrm{c}^\dagger$), and dephasing ($\hat{n}_\mathrm{c}$) errors. Here, we consider a generalized error set:
\begin{eqnarray}
\bar{\mathcal{E}} = \{\hat{a}_\mathrm{c}, \hat{a}_\mathrm{c}^2,  ..., \hat{a}_\mathrm{c}^L, \hat{a}_\mathrm{c}^\dagger, \hat{a}_\mathrm{c}^{\dagger 2}, ..., \hat{a}_\mathrm{c}^{\dagger G}, \hat{n}_\mathrm{c}, \hat{n}_\mathrm{c}^2, ...,\hat{n}_\mathrm{c}^D\},
\end{eqnarray}
encompassing $L$-photon loss, $G$-photon gain, and $D$-dephasing error events. These errors can be corrected using generalized binomial codes~\cite{michael2016} with spacing $S = L + G$ and maximum order $N = \max\{L, G, 2D\}$:
\begin{eqnarray}
\left\vert 0_\mathrm{L}\right\rangle &=& \frac{1}{\sqrt{2^N}}\sum_{p, \mathrm{even}}^{[0,N+1]}\sqrt{C_{N+1}^p} \left\vert p(S+1)\right\rangle,\notag \\ 
\left\vert 1_\mathrm{L}\right\rangle &=& \frac{1}{\sqrt{2^N}}\sum_{p, \mathrm{odd}}^{[0,N+1]}\sqrt{C_{N+1}^p} \left\vert p(S+1)\right\rangle,
\end{eqnarray}
where $C_{N+1}^p$ represents the binomial coefficients. We define corresponding error subspaces: $\mathcal{E}_k^{\downarrow}$ for $k$-photon loss, $\mathcal{E}_k^{\uparrow}$ for $k$-photon gain, and $\mathcal{E}_k^{\phi}$ for $k$-th order dephasing error. The cascaded AQEC scheme, illustrated in Fig.~\ref{fig:Fig_cAQEC}, implements a single unitary operation $\hat{U}_\mathrm{AQEC}$ that performs hierarchical correction across all three error categories simultaneously. For dominant photon loss errors, this operation corrects states in $\mathcal{E}_n^{\downarrow}$ to $\mathcal{E}_{n-1}^{\downarrow}$, with final correction from $\mathcal{E}_1^{\downarrow}$ to the code space $\mathcal{C}$, while flipping the ancilla qubit to $|e\rangle$. Similarly, for photon gain errors, states in $\mathcal{E}_n^{\uparrow}$ are corrected to $\mathcal{E}_{n-1}^{\uparrow}$, and for dephasing errors, states in $\mathcal{E}_n^{\phi}$ are corrected to $\mathcal{E}_{n-1}^{\phi}$, with each correction step accompanied by a qubit flip. However, when photon gain errors reach $\mathcal{E}_1^{\uparrow}$ or dephasing errors reach $\mathcal{E}_1^{\phi}$, the states remain in these error subspaces while the ancilla qubit stays in the $|g\rangle$ state. The ancilla qubit is subsequently reset to $|g\rangle$ after each correction cycle.

The modular design of our cascaded AQEC scheme enables simultaneous correction of photon loss, photon gain, and dephasing errors within a single QEC cycle, providing comprehensive protection against multiple error mechanisms.

\section{Quantum-enhanced metrology via AQEC}
\subsection{Theory}
A quantum metrology experiment typically consists of three consecutive steps: preparation of a probe state $\hat{\rho}_0$, imprinting an unknown parameter $\delta$ onto the probe state through operation $\hat{U}(\delta)=e^{-i\delta \hat{G}}$, i.e. $\hat{\rho}(\delta)=\hat{U}(\delta)\hat{\rho}_0\hat{U}^\dagger(\delta)$, and finally performing a measurement $\hat{M}$ on the evolved state to extract information about the estimated parameter. The precision of estimating the unknown parameter can be quantified by the Fisher information $\mathcal{F}$ according to the Cram\'er-Rao bound~\cite{braunstein1994}, that is $1/(\Delta\delta)^2\le\mathcal{F}$ for an unbiased estimator. Here, $\hat{G}$ is a process generator. The Fisher information can be extracted from the probability distribution of $P(m|\delta)$, conditioned on $\delta$, as expressed by the formula~\cite{braunstein1994}:
\begin{eqnarray}
\mathcal{F}=\sum_m\frac{1}{P(m|\delta)}(\frac{\partial P(m|\delta)}{\partial \delta})^2.
\end{eqnarray}

However, the parameter imprinting process is susceptible to environmental noise, which perturbs the probe state and degrades the detection signal, ultimately elevating estimation uncertainty. This challenge is exacerbated in time-dependent measurement protocols where extended integration periods improve precision at the cost of higher error rates. Quantum error correction (QEC) provides a viable pathway to resolve this fundamental trade-off. This work focuses on the estimation of a slight frequency shift, utilizing the Fock superposition state $(|m\rangle+e^{i\varphi}|n\rangle)/\sqrt{2}$ ~\cite{wang2022NC} with initial variable phase $\varphi$ as the probe state, where $m$ and $n$ are positive integers. The frequency shift parameter $\delta$ can then be imprinted onto this probe state through a unitary evolution $\hat{U}(\delta,\tau_{\mathrm{int}})=e^{-i\delta \tau_{\mathrm{int}}\hat{a}^\dagger\hat{a}}$, resulting in the state $(|m\rangle+e^{i[(n-m)\delta\tau_\mathrm{int}+\varphi]}|n\rangle)/\sqrt{2}$, where $\tau_{\mathrm{int}}$ presents an interrogation time. During the interrogation time, the dominant single-photon-loss error would transform the initial probe state to an error state $(\sqrt{m}|m-1\rangle+\sqrt{n}e^{i[(n-m)\delta\tau_\mathrm{int}+\varphi]}|n-1\rangle)/\sqrt{m^2+n^2}$, whereas the imprinted parameter information remains undamaged. Therefore, QEC techniques can be applied to revert the error space $\{|m-1\rangle,|n-1\rangle\}$ back to the correct space $\{|m\rangle,|n\rangle\}$, after which parameter estimation measurement proceeds. It is noteworthy that a perfect quantum error-correcting code cannot be designed for this superposition state $(|m\rangle+e^{i\varphi}|n\rangle)/\sqrt{2}$ due to violating the Knill-Laflamme (KL) conditions~\cite{knill1997}, particularly the conditions for perfectly correcting single-photon loss. Nevertheless, approximate QEC techniques can still be employed to mitigate single-photon loss errors during sensing.

For parameter estimation measurements, we employ the quantum optimal control technique (using GRAPE algorithm, see Sec.~\ref{sec:grape}) to implement projective measurements in a set $\{|+\rangle\langle+|, |-\rangle\langle-|\}$ on the evolved quantum state, where $|\pm\rangle=(|m\rangle\pm|n\rangle)/\sqrt{2}$ are the optimal measurement basis for phase estimation~\cite{wang2022NC}. The measurement outcomes are then mapped onto the population distribution between the ground and excited states of the ancilla qubit, which exhibit period oscillations:
\begin{eqnarray}
P_g&=&P(+|\delta)=A\cos\left(\delta(n-m)\tau_{\mathrm{int}}+\varphi\right)+B,\notag\\
P_e&=&P(-|\delta)=1-P_g,
\label{eqn:pg_cos}
\end{eqnarray}
where $A$ is the oscillation amplitude and $B$ is the offset. The corresponding Fisher information is then calculated as
\begin{eqnarray}
\mathcal{F}=\max_\varphi\left(\sum_m\frac{1}{P(m|\delta)}(\frac{\partial P(m|\delta)}{\partial \delta})^2\right)=\frac{A^2(n-m)^2\tau_{\mathrm{int}}^{2}}{B(1-B)}.
\end{eqnarray}

\subsection{Experimental details}
\begin{figure*}[tb]
    \includegraphics{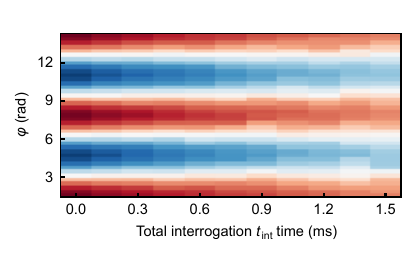}
    \caption{Measured ancilla ground-state population oscillating as a function of the initial phase $\varphi$ with different total interrogation times $t_\mathrm{int}=M\tau_\mathrm{int}$ for the probe state $(|0\rangle+e^{i\varphi}|1\rangle)/\sqrt{2}$. 
}
    \label{fig:Fig_01sensingdata}
\end{figure*}

This work demonstrates the estimation of a slight frequency shift $\delta$ using the Fock superposition state $(|1\rangle+|4\rangle)/\sqrt{2}$ as the probe state, protected by autonomous approximate QEC technology. The detailed analysis and results of AQEC for protecting the probe state $(|1\rangle+e^{i\varphi}|4\rangle)/\sqrt{2}$ against single-photon loss are presented in Sec.~\ref{sec:AQEC_other}. By experimentally varying the phase $\varphi$ through a virtual phase operation to emulate phase accumulation from a frequency shift, we obtain a $\varphi$-dependent oscillation curve. Using an interrogation time of $\tau_{\mathrm{int}}=150~\mu$s per cycle, we repeatedly apply the AQEC process over $M$ cycles to extend the total phase accumulation time to $t_{\mathrm{int}}=M\tau_\mathrm{int}$. The measured Ramsey oscillations without and with AQEC protection are presented in Figs.~4b-c of the main text, while the results for uncorrected probe states $(|1\rangle+e^{i\varphi}|4\rangle)/\sqrt{2}$ are presented in Fig.~\ref{fig:Fig_01sensingdata} for comparison. 
These Ramsey oscillations are fitted to the function in Eq.~\ref{eqn:pg_cos}. The extracted fitting parameters $A$ and $B$ are used to calculate the corresponding normalized Fisher information by $F=\mathcal{F}/t_\mathrm{tot}$. Here $t_\mathrm{tot}$ represents the total single-shot measurement time, including probe state initialization, signal interrogation, AQEC operation, characterization, and readout times. The resulting normalized Fisher information versus the total interrogation time for these three cases is presented in Fig.~4d of the main text. 

Here we provide detailed calculations of the normalized Fisher information at the optimal point for each curve in Fig.~4d of the main text. For the probe state $(|1\rangle+|4\rangle)/\sqrt{2}$ under AQEC protection, the optimal performance with maximal normalized Fisher information occurs after three AQEC cycles. Sinusoidal fitting to the corresponding Ramsey oscillation yields fitting parameters $A = 0.296$ and $B = 0.516$. Thus, the normalized Fisher information of the corrected probe state $(|1\rangle+|4\rangle)/\sqrt{2}$ is then calculated as
\begin{eqnarray}
F_{1,4}^c&=&\frac{A^2}{B(1-B)}\times\frac{(n-m)^2t_{\mathrm{int}}^{2}}{t_\mathrm{tot}}\notag\\&=&0.351\times\frac{(4-1)^2\,(150~\mathrm{\mu s} \times 3)^{2}}{(150+2.71)~\mathrm{\mu s} \times 3 +6.08~\mathrm{\mu s}}=1.38 \mathrm{kHz}^{-1},
\end{eqnarray}
where the total interrogation time for three cycles is $t_{\mathrm{int}}=150~\mathrm{\mu s}\times 3$, and the total experimental time $t_{\mathrm{tot}}=(150~\mathrm{\mu s}+2.71~\mathrm{\mu s})\times 3+6.08~\mathrm{\mu s}$ includes interrogation ($150~\mathrm{\mu s}$), correction pulse ($2~\mathrm{\mu s}$), and qubit reset ($710~\mathrm{ns}$) for three times, and fixed overheads for encoding-decoding operation ($4.08~\mathrm{\mu s}$) and readout ($2~\mathrm{\mu s}$).

For the uncorrected probe state $(|1\rangle+|4\rangle)/\sqrt{2}$, the corresponding normalized Fisher information is calculated as
\begin{eqnarray}
F_{1,4}=0.242\times \frac{(4-1)^2\,(150~ \mathrm{\mu s} \times 2)^{2}}{150~ \mathrm{\mu s}\times 2 +6.08~\mathrm{\mu s}}=0.644 \mathrm{kHz}^{-1},
\end{eqnarray}
with fitting parameters $B=0.482$ and $A=0.246$ from the third experimental trace in Fig.~4b in the main text. 

And for the uncorrected probe state $(|0\rangle+|1\rangle)/\sqrt{2}$, the corresponding normalized Fisher information is
\begin{eqnarray}
F_{0,1}=0.273\times \frac{(150~\mathrm{\mu s} \times 8)^{\,2}}{150~ \mathrm{\mu s}\times 8 +5.08~\mathrm{\mu s}}=0.327 \mathrm{kHz}^{-1},
\end{eqnarray}
where the fixed overhead is $5.08~\mathrm{\mu s}$ (including $3.08~\mathrm{\mu s}$ for encoding-decoding operation and $2~\mathrm{\mu s}$ for readout), and the fitting parameters are $B=0.525$ and $A=0.261$.

Although the sensing Hamiltonian $\hat{H}_\mathrm{int}=\delta\hat{a}^\dagger\hat{a}$ does not fully satisfy the Hamiltonian-not-in-Lindblad span (HNLS) condition~\cite{zhou2018,wang2022NC}, i.e., it is not fully orthogonal to the cavity noise set $\{\hat{I},\hat{a}^\dagger,\hat{a},\hat{a}^\dagger\hat{a}\}$, thus preventing to achieve the Heisenberg limit, our quantum metrology scheme still benefits from QEC with enhanced sensitivity. Specifically, the corrected probe state $(|1\rangle+|4\rangle)/\sqrt{2}$ yields a metrological gain of $10\log_{10}F_{1,4}^c/F_{1,4}=3.3$~dB in the normalized Fisher information compared to the uncorrected probe state, and a gain $10\log_{10}F_{1,4}^c/F_{0,1}=6.3$~dB over the probe state $(|0\rangle+|1\rangle)/\sqrt{2}$, demonstrating the significant enhancement through AQEC protection.

%